\documentclass[paper]{JHEP3}
\usepackage{cite}
\usepackage{amssymb,amsmath}
\usepackage{graphicx}
\usepackage{booktabs}
\usepackage{subfigure}
\usepackage{enumerate}

\title{
Double real radiation corrections to $t\bar{t}$ production at the LHC: the $gg\rightarrow t\bar{t}q\bar{q}$ channel}

\author{
G.~Abelof, A.~Gehrmann--De Ridder\\
Institute for Theoretical Physics, ETH, CH-8093 Z\"urich, Switzerland}


\abstract{
We present the double real radiation contributions to the $t \bar t$ hadronic production cross section stemming from the partonic process $gg\rightarrow t\bar{t}q\bar{q}$. We explicitly construct the antenna subtraction terms for this gluon-gluon initiated process starting from the double soft behaviour of the double real radiation matrix elements using soft currents. Those subtraction terms, given in leading and subleading colour contributions, require the use of new genuine NNLO four-parton antenna functions involving massive fermions. Those are also presented together with their infrared limits in this paper. We checked the validity of our subtraction terms numerically by showing that the ratio between the real radiation matrix elements and the subtraction terms approaches unity in all single and double unresolved regions of phase space.}

\allowdisplaybreaks[4]

\newcommand{\beq}{\begin{equation}}
\newcommand{\eeq}{\end{equation}}
\newcommand{\beqa}{\begin{eqnarray}}
\newcommand{\eeqa}{\end{eqnarray}}
\newcommand{\cm}{{\cal M}^0}
\newcommand{\ds}{{\rm d}\hat{\sigma}}
\newcommand{\dphi}{{\rm d}\Phi}
\newcommand{\wt}{\widetilde}
\newcommand{\re}{{\rm{Re}}}
\newcommand{\norm}{{\cal N}}
\newcommand{\ssoft}[3]{{\cal S}_{#1 #2 #3}}
\newcommand{\soft}[4]{{\cal S}_{#1 #2 #3 #4}}
\newcommand{\order}[1]{{\cal O}(#1)}
\newcommand{\q}[1]{#1_q}
\newcommand{\qb}[1]{#1_{\bar{q}}}

\newcommand{\Q}[1]{#1_Q}
\newcommand{\Qb}[1]{#1_{\bar{Q}}}
\newcommand{\qi}[1]{\hat{#1}_q}
\newcommand{\qbi}[1]{\hat{#1}_{\bar{q}}}

\newcommand{\gl}[1]{#1_g}
\newcommand{\gli}[1]{\hat{#1}_g}
\newcommand{\ph}[1]{#1_{\gamma}}
\newcommand{\phin}[1]{\hat{#1}_{\gamma}}
\newcommand{\del}[2]{\delta_{i_{#1},i_{#2}}}

\def\JET{J}
\def\e{\epsilon}

\begin{document}
\bibliographystyle{JHEP-2}

\section{Introduction}\label{sec.intro}
With a mass $m_{t}=173 \pm 1.3$  GeV, the top quark is the heaviest quark produced at colliders and due to its narrow width, it decays before it hadronises. Since its discovery at the Fermilab Tevatron~\cite{Abachi:1995iq,Abe:1995hr}, a number of its properties (mass, couplings) have been determined to an accuracy of ten to twenty per cent. With the large number of top quark pairs expected to be produced at the LHC \cite{Silva:2012di}, more accurate experimental measurements will become available, with some observables such as the $t\bar{t}$ production cross section, being measured with an accuracy of the order of five percent. These very precise measurements have to be matched unto equally accurate theoretical predictions. Therefore, fixed-order calculations of these observables need to be considered at least at next-to-leading (NLO), if not even at next-to-next-to-leading order (NNLO) in perturbative QCD. 

NLO predictions for top quark pair production cross sections in the narrow width approximation have been known already for some time~\cite{Nason:1989zy}, and more recently off-shell effects and spin correlations between the production and the decay of the top quarks have been included in~\cite{Bevilacqua:2010qb,Denner:2010jp,Melnikov:2011qx,Denner:2012yc}. The next-to-leading-logarithmic resummation (NLL) are also known~\cite{Cacciari:2008zb,Kidonakis:2008mu,Moch:2008qy}, and even the NNLL resummation effects have been completed in~\cite{Ahrens:2010zv}. These predictions combined yield a theoretical uncertainty of the order of ten per cent.  The same precision is available for single top quark production~\cite{Harris:2002md}, top-pair-plus-jets production~\cite{Dittmaier:2007wz,Bevilacqua:2010ve,Melnikov:2010iu} and for top-pair-plus-bottom-pair production~\cite{Bevilacqua:2009zn,Bredenstein:2010rs}.

At NNLO, an increasing number of intermediate results have become available recently. Most notably, the inclusive total hadronic $t \bar{t}$ production cross section induced by the all-fermion partonic processes has been computed~\cite{Baernreuther:2012ws,Czakon:2012zr}. NNLO calculations involving massive quarks require the same ingredients as their massless counterparts, namely, double real ${\rm d}\sigma^{RR}$, mixed real-virtual, ${\rm d}\sigma^{RV}$, and double virtual contributions ${\rm d}\sigma^{VV}$. For top pair production, the double virtual corrections built with one-loop amplitudes squared have been fully evaluated in \cite{Anastasiou:2008vd,Kniehl:2008fd,Korner:2008bn}. Although the full two-loop matrix elements are not yet available, there is on one hand a purely numerical evaluation of the $q\bar q \to t\bar{t}$ channel~\cite{Czakon:2008zk}, and on the other hand analytical results in both the quark-antiquark and gluon-gluon initiated channels have been obtained in~\cite{Bonciani:2008az,Bonciani:2009nb,Bonciani:2010mn}. The infrared structure of the one-loop amplitudes for the mixed real-virtual contributions has been studied in~\cite{Bierenbaum:2011gg}.

While infrared singularities from purely virtual corrections are obtained immediately after integration over the loop momenta, their extraction is more involved for real emission (or mixed real-virtual) contributions. There, the infrared singularities only become explicit after integrating  the matrix elements over the phase space appropriate to the differential cross section under consideration. Since hadronic observables depend in general in a non trivial manner on the experimental criteria used to define them, they can only be calculated numerically. The computation of hadronic observables including higher order corrections therefore requires a systematic procedure to cancel infrared singularities among different partonic channels before any numerical computation of the observable can be performed. 

Subtraction methods explicitly constructing infrared subtraction terms which coincide with the full matrix element in the unresolved limits and are sufficiently simple to be integrated analytically are well-known solutions to this problem. Starting from methods for subtraction at NLO~\cite{Campbell:1998nn,Catani:1996vz,Kosower:1997zr,Kunszt:1992tn,Somogyi:2009ri}, several NNLO subtraction methods have been proposed in the literature~\cite{Aglietti:2008fe,Bierenbaum:2011gg,Bolzoni:2009ye,Bolzoni:2010bt,Catani:2007vq,Czakon:2010td,Czakon:2011ve,Frixione:2004is,Kilgore:2004ty,Kosower:2002su,Somogyi:2005xz,Somogyi:2006db,Somogyi:2008fc,Weinzierl:2003fx, GehrmannDeRidder:2005cm,GehrmannDeRidder:2011aa,Glover:2010im}, and have been worked out to a varying level of sophistication. Numerical methods not directly based on subtraction have also been proposed \cite{Binoth:2000ps,Binoth:2003ak,Heinrich:2008si,Heinrich:2002rc,Anastasiou:2003gr, Binoth:2004jv, Heinrich:2006ku,Carter:2010hi,Anastasiou:2004xq,Anastasiou:2007mz,Anastasiou:2010pw,Buehler:2012cu,Melnikov:2006kv}.

Employing a subtraction method, the NNLO partonic cross section for a hadronic observable can be written as~\cite{Glover:2010im,GehrmannDeRidder:2011aa}
\beqa
{\rm d}\hat\sigma_{NNLO}&=&\int_{{\rm{d}}\Phi_{m+2}}\left({\rm{d}}\hat\sigma_{NNLO}^{RR}-{\rm{d}}\hat\sigma_{NNLO}^S\right)
+\int_{{\rm{d}}\Phi_{m+2}}{\rm{d}}\hat\sigma_{NNLO}^S\nonumber\\
&+&\int_{{\rm{d}}\Phi_{m+1}}\left({\rm{d}}\hat\sigma_{NNLO}^{RV}-{\rm{d}}\hat\sigma_{NNLO}^{V S}\right)
+\int_{{\rm{d}}\Phi_{m+1}}{\rm{d}}\hat\sigma_{NNLO}^{V S}
+\int_{{\rm{d}}\Phi_{m+1}}{\rm{d}}\hat\sigma_{NNLO}^{MF,1}\nonumber\\
&+&\int_{{\rm{d}}\Phi_m}{\rm{d}}\hat\sigma_{NNLO}^{VV}
+\int_{{\rm{d}}\Phi_m}{\rm{d}}\hat\sigma_{NNLO}^{MF,2}\label{eq.sigNNLO}.
\eeqa
In this equation, ${\rm d} \hat\sigma^{S}_{NNLO}$ denotes the subtraction term for the $(m+2)$-parton final state which behaves like the double real radiation contribution ${\rm d} \hat\sigma^{RR}_{NNLO}$ in all singular limits. Likewise, ${\rm d} \hat\sigma^{VS}_{NNLO}$ is the one-loop virtual subtraction term coinciding with the one-loop $(m+1)$-final state ${\rm d} \hat\sigma^{RV}_{NNLO}$ in all singular limits. The two-loop correction to the $(m+2)$-parton final state is denoted by ${\rm d}\hat\sigma^{VV}_{NNLO}$.  In addition, as there are partons in the initial state, two mass factorisation contributions, ${\rm d}\hat\sigma^{MF,1}_{NNLO}$ and ${\rm d}\hat\sigma^{MF,2}_{NNLO}$, for the $(m+1)$- and $m$-particle final states respectively, need to be taken into account.

For QCD observables involving massive final state particles, fewer subtractions terms are required since the real matrix elements develop singular behaviours in fewer regions of phase space than in the massless case. Indeed, for those observables, QCD radiation from massive particles can lead to soft divergencies but cannot lead to strict collinear divergencies, since the mass acts as an infrared regulator. The quasi-collinear logarithms \cite{Abelof:2011jv,Catani:2000ef,Catani:2002hc,GehrmannDeRidder:2009fz} of the form $\log(m_t^2/s_{ij})$ are not enhanced in the kinematical configuration under consideration, and we shall therefore ignore them, as it is also done in \cite{Baernreuther:2012ws,Czakon:2012zr}.

At NLO, the computation of hadronic observables can be performed using an extension of the dipole formalism presented in \cite{Catani:2000ef,Catani:2002hc}. Alternatively, the antenna subtraction formalism extended to include the treatment of observables involving massive fermions can also be used \cite{Abelof:2011jv,GehrmannDeRidder:2009fz}. 

Within the antenna subtraction formalism, originally developed for the computation of jet observables in $e^+e^-$ annihilation \cite{GehrmannDeRidder:2005cm} and which uses colour ordering properties of amplitudes in a crucial manner, the subtraction terms are constructed using the fundamental factorisation properties of QCD amplitudes and phase spaces in their collinear and soft limits. These terms rely on the following essential ingredients: 

\begin{itemize}
\item[(1)] A set of antenna functions of various types, which capture all unresolved radiation emitted between two hard partons, the radiators. These can be either massless or massive and depending on where the two radiators are located, in the initial or in the final state, we distinguish three types of antennae: final-final, initial-final and initial-initial. 

\item[(2)] An exact momentum conserving and Lorentz invariant phase space factorisation based on $3\to 2$ and $4\to 2$ mappings between on-shell partons in all three configurations.  

\item[(3)] Phase-space mappings defining the momenta present in the factorised form of the matrix elements in terms of the original momenta present in the real radiation matrix elements for which the subtraction terms are constructed. 
\end{itemize}

The subtraction terms are then built with products of antenna functions and reduced matrix elements squared in all three configurations: final-final, initial-final and initial-initial.
The framework for the construction of NNLO antenna subtraction terms for hadronic jet observables (involving massless partons) has been set up in~\cite{GehrmannDeRidder:2011aa,Glover:2010im} in the context of a proof-of-principle implementation of the contribution of the purely gluonic contributions to di-jet production at hadron colliders.

Compared to the massless antenna formalism, the presence of massive partons in the final states modifies the subtraction terms in a non-trivial way. Parton masses lead to modified kinematics and have to be taken into account for the phase space factorisations~\cite{Abelof:2011ap}. Furthermore, non-vanishing masses also modify the soft behaviour of the real radiation matrix elements as will be explained in section \ref{sec.subterms}.

In~\cite{GehrmannDeRidder:2009fz,Abelof:2011jv}, the antenna subtraction method was extended at the NLO level and employed to compute the real NLO corrections to the $t \bar{t}$ and $t \bar{t}+jet$  hadronic production cross section. In~\cite{Abelof:2011ap} we extended the method to the NNLO level, and derived the double real emission corrections to $t\bar{t}$ hadronic production cross section coming from all partonic channels involving fermions only. The aim of this paper is to show how, within the framework presented in~\cite{Abelof:2011ap}, partonic processes involving gluons can also be evaluated. We shall employ this formalism to evaluate the double real contributions to the hadronic top quark pair production arising from the partonic process $gg \to Q \bar{Q} q \bar{q}$. Although the framework established in ~\cite{Abelof:2011ap} can be used, new essential ingredients are required here, as the infrared structure of the partonic process dealt with in this paper differs significantly from the infrared behaviour of the all-fermion processes treated in \cite{Abelof:2011ap}.

The paper is organised as follows: In section 2, we briefly present the infrared structure of double real contributions of jet observables at hadron colliders. In section \ref{sec.antennae} we give a list of all genuine new NNLO massive four-parton tree-level antennae required in our calculation while in section \ref{sec.limits} we list the behaviour of these antennae in their infrared limits. In section \ref{sec.subterms}, we first present the double soft behaviour of the real matrix element associated to the process $gg \to Q \bar{Q} q \bar{q}$. This enables us to construct the subtraction term capturing the double unresolved features of the double real matrix elements explicitly. All subtraction terms required to capture the single and double unresolved behaviour of the double real matrix elements in leading and subleading colour will be given in that section too. In section \ref{sec.results}, we test the validity of the subtraction terms. We check that the ratio between the real radiation matrix elements and the corresponding subtraction terms approaches unity in all single and double unresolved limits. Finally, section \ref{sec.conclusions} contains our conclusions and an outline. Three appendices are also enclosed: appendix A includes a list of the single unresolved factors needed in the context of our calculation, appendix B contains a list of all tree-level three-parton antennae together with their infrared limits, while the known four-parton antennae that are used in the construction of our subtraction terms are given in appendix C.

\section{Double real radiation contributions to heavy quark pair production}\label{sec.formalism} 
The double real emission contributions to $p p \to Q\bar{Q}+(m-2){\rm jets}$ at the partonic level read 
 \begin{eqnarray}
\lefteqn{{\rm d}\hat\sigma^{RR}_{NNLO}(p_1,p_2)=
{\cal N}_{NNLO}^{RR}\,
\sum_{{m}}{\rm d}\Phi_{m+2}(p_{Q},p_{\bar{Q}},p_{5},\ldots,\,p_{m};
p_1,p_2) }\nonumber \\ && \times 
\frac{1}{S_{{m}}}\,
|{\cal M}_{m+4}(p_{Q},p_{\bar{Q}},p_{5},\ldots,p_{m};p_1,p_2)|^{2}\; 
\JET_{m}^{(m+2)}(p_{Q},p_{\bar{Q}},p_{5},\ldots,p_{m}).\hspace{3mm} \\
&\equiv&
{\cal N}_{NNLO}^{RR}\,
\sum_{{m}}{\rm d}\Phi_{m+2}(p_{3}, \ldots, p_{m+4}; p_1,p_2) \nonumber \\ && \times 
\frac{1}{S_{{m}}}\,
|{\cal M}_{m+4}(p_{3},\ldots, p_{m+4},;p_1,p_2)|^{2}\; 
\JET_{m}^{(m+2)}(p_{3},\ldots,p_{m+4}),\hspace{3mm}\label{eq.real}
\end{eqnarray}
where $p_1$ and $p_{2}$ are the momenta of the initial state partons and 
where the last line is obtained by relabelling all final state partons.
In eq.(\ref{eq.real}), $S_{m}$ is a symmetry factor for identical massless partons in the final state while the jet function denoted by $\JET_{m}^{(m+2)}$ ensures that out of $m$ massless partons and a $Q\bar{Q}$ pair, an observable with a pair of heavy quark jets in addition to $(m-2)$ jets, is built. The NNLO normalisation factor ${\cal N}^{RR}_{NNLO}$ includes all QCD-independent factors as well as the dependence on the renormalised QCD coupling constant $\alpha_s$. It is related to the normalisation factor present at leading order, ${\cal N}_{LO}$, which depends on the specific process and parton channel under consideration. For the process under consideration in this paper, this relation will be specified in section~\ref{sec.subterms}.
$\sum_{m}$ denotes the sum over all configurations with $m$ massless partons while  ${\rm d}\Phi_{m+2}$ is the phase space for an $m+2$-parton final state containing $m$ massless and two massive partons with total four-momentum $p_1^{\mu}+p_2^{\mu}$.

In eq.(\ref{eq.real}), $|{\cal M}_{m+4}|^2$ denotes a colour-ordered tree-level matrix element squared with $m+2$ final state partons, out of which two are massive and two are initial state partons. These terms only account for the leading colour contributions to the squared matrix elements, since subleading colour contributions involve in general interferences between sub-amplitudes with different colour orderings. However, to keep the notation simpler we denote these interference terms also as $|{\cal M}_{m+4}|^2$. 

The NNLO contribution given in eq.(\ref{eq.real}) develops singularities if one or two final state partons are unresolved (soft or collinear). Depending on the colour connection between these unresolved partons, the following configurations must be distinguished~\cite{GehrmannDeRidder:2005cm,Glover:2010im}

\begin{itemize}
\item[(a)] One unresolved parton but the experimental observable selects only $m$ jets.
\item[(b)] Two colour-connected unresolved partons (colour-connected).
\item[(c)] Two unresolved partons that are not colour-connected but share a common radiator (almost colour-unconnected).
\item[(d)] Two unresolved partons that are well separated from each other in the colour chain (colour-unconnected).
\item[(e)] Compensation terms for the over-subtraction of large angle soft emission.
\end{itemize}
This separation among subtraction contributions according to colour connection is valid in all final-final, initial-final or initial-initial configurations and, in any of them, the subtraction formulae have a characteristic structure in terms of the required antenna functions. This antenna structure has been derived for processes involving only massless partons, for the final-final and initial-final cases in \cite{Daleo:2009yj,GehrmannDeRidder:2005cm} and \cite{Glover:2010im} for the initial-initial case. Note that the presence of massive partons in the final state does not modify the general structure of the subtraction terms required to match the unresolved features given above as explained in \cite{Abelof:2011ap}.    

For the partonic process $gg \to Q \bar{Q} q \bar{q}$ considered in this paper, the configuration $(c)$, the almost colour-unconnected case, and the configuration $(e)$, regarding the treatment of large angle soft radiation, do not occur. On one hand, the colour structure of the amplitude for this process does not allow configuration $(c)$ and, on the other hand, the absence of final state gluons forbids configuration $(e)$. The discussion of these configurations will therefore be treated elsewhere. Furthermore, the general structure of the subtraction terms required in configurations $(a)$, $(b)$ and $(d)$ has been presented in great detail in \cite{Abelof:2011ap}. Since for the treatment of the double real emission contributions to the hadronic $t \bar{t}$ production stemming from the partonic process $ gg \to Q \bar{Q} q \bar{q}$, this structure remains unchanged, it will not be discussed further here either. 

However, although the general structure of the subtraction terms constructed for the all fermion processes and presented in \cite{Abelof:2011ap} and the one of the partonic process dealt with in this paper are comparable, the essential ingredients required to build these subtraction terms are different. The antenna functions capturing the infrared behaviour of the real matrix element and which depend on the nature of the partons involved in the process under consideration are different. The new genuine NNLO four parton massive antennae required to construct the subtraction term for the process  $gg\rightarrow Q\bar{Q}q\bar{q}$ and presented in section \ref{sec.subterms}, will be derived in section \ref{sec.antennae} while  their infrared limits will be given in section \ref{sec.limits}.

\section{Antenna functions}\label{sec.antennae}
Antenna functions are the key ingredients needed to build subtraction terms in the antenna subtraction method and they can also be used as evolution kernels in parton showers~\cite{GehrmannDeRidder:2011dm,Giele:2007di,Giele:2011cb,LopezVillarejo:2011ap}. The general features of the different types of antenna functions needed for the double real corrections to $t\bar{t}$ production have been already discussed in \cite{Abelof:2011ap}. We shall here just recall that they are calculated as ratios of physical colour-ordered matrix elements squared, they can be categorised according to the parton flavour that they collapse onto in their singular limits, and according to whether the hard radiators are in the initial or in the final state. While at NLO only three-parton tree-level antennae are needed, NNLO subtraction terms in addition require three-parton one-loop antennae and four-parton tree-level antennae. The one-loop antenna, however, are not needed in the treatment of double real radiation corrections.

For the subtraction terms that will be presented in section \ref{sec.subterms} for the partonic process $gg \rightarrow Q\bar{Q}q\bar{q}$, the three-parton antennae that are needed are of A, D, E and F-type in different configurations with massive and/or massless partons. All these antennae have been computed and integrated in~\cite{Abelof:2011jv,Daleo:2006xa,GehrmannDeRidder:2005cm,GehrmannDeRidder:2009fz}, with the exception of one flavour-violating A-type antenna. This antenna function, which involves a gluon in the initial state, a massive quark and a massless antiquark in the final state will be presented together with its integrated form and infrared limits in appendix~\ref{sec.3partonantennae}. The unintegrated form of all other required three-parton antennae will also be presented there for completeness.

The genuine NNLO four-parton antennae that are needed for the partonic process under consideration are of B, E and G-type. The B and G-type antennae are known and will be given in the appendix~\ref{sec.4partonantennae}. The E type antennae are new and shall be derived below.

For the labelling of the partons in the antenna functions used throughout this paper we shall use the same conventions as in \cite{Abelof:2011ap}: Massless quarks will be indexed with $q$ while massive ones with $Q$ and their mass with $m_Q$. Partons crossed to the initial state are denoted with a hat. The first and the last particles in the argument of a given antenna are the hard radiators, while the partons placed in between the radiators are the unresolved particles. In order to make the mass-dependence in the expressions of the antenna functions explicit, we define our invariants as $s_{ij}=2 p_{i} \cdot p_{j}$. Finally, for conciseness, the $ {\cal O}(\epsilon)$ pieces of the antenna functions will be omitted.

The B-type antenna required to define our subtraction terms is massive and is only needed in its final-final form. It is obtained from the ratio of the matrix elements squared for the physical processes $\gamma^{*}\to Q\bar{Q}q\bar{q}$ and $\gamma^{*} \to Q \bar{Q}$, and it is used to subtract the infrared singularities associated to the emission of an unresolved $q\bar{q}$ pair between the massive $Q\bar{Q}$ radiator pair. This final-final massive B-type antenna is known in unintegrated and integrated form~\cite{Bernreuther:2011jt}, and will be given in the appendix for completeness. Its infrared limits will be recalled in section \ref{sec.limits}.

We also use a massless four-parton G-type antenna in its initial-initial form. The expression of this antenna is obtained by crossing two gluons in the corresponding final-final massless G-type antenna, which has been defined in \cite{GehrmannDeRidder:2005cm} as the ratio of the processes $H\to ggq\bar{q}$ and $H\to gg$. This antenna is needed to subtract the infrared singularities associated to the emission of a massless $q\bar{q}$ pair radiated between the initial state gluons. Its unintegrated form will be given in the appendix while its infrared limits, which have not been documented so far, will be presented in section \ref{sec.limits}. Note as well, that the integrated form of this antenna has just been derived in~\cite{GehrmannDeRidder:2012ja}.

The new massive four-parton E-type antenna functions are needed in their initial-final form. They can be obtained by crossing the gluon in the corresponding final-final E-type antennae, which are derived from the ratio of the processes $\tilde{\chi}\to \tilde{g}gq\bar{q}$ and $\tilde{\chi}\to \tilde{g}g$ with the massive gluino $\tilde{g}$ playing the role of the massive (anti) quark of mass $m_Q$. The full amplitude for the process $\tilde{\chi}\to \tilde{g}gq\bar{q}$ contains leading and subleading colour pieces~\cite{GehrmannDeRidder:2005aw}. By squaring the leading colour piece, in which the $q\bar{q}$ is emitted between the gluino and the gluon in the colour chain, the $E_4^0$ antenna is obtained, while, by squaring the subleading colour piece, in which the gluon is emitted between the $q\bar{q}$ pair, the $\wt{E}_4^0$ antenna is obtained. 

The infrared limits of these E-type antennae derived below will be given in section~\ref{sec.limits}. Note also that, there is on-going work~\cite{Abelof:InPreparation} towards the integration of these initial-final massive antennae.

To account for the infrared limits associated to the emission of an unresolved $q\bar{q}$ pair between a massive (anti) quark and an initial state gluon we use the following antenna function,
\beqa
E_4^0(\Q{1},\q{3},\qb{4},\gli{2})&=&\frac{1}{\left(Q^2+m_Q^2\right)^2}\bigg\{
\frac{s_{13} s_{14}} {s_{12} s_{234}}  + \frac{s_{14}}{ s_{12} s_{34} s_{24}}  \left[- s_{13} s_{23}+ s_{13}^2 + s_{14}^2 \right]\nonumber \\ 
&& + \frac{s_{14}}{ s_{12} s_{34} s_{134}}  \left[ 2 s_{14} s_{23}+ 2 s_{14} s_{24}+ s_{23}^2- s_{24}^2\right]\nonumber \\ 
&&+ \frac{s_{14}}{ s_{12} s_{24} s_{234}}  \left[-s_{13} s_{23}+ s_{13}^2+ s_{14}^2\right]+ \frac{s_{14}}{ s_{12} s_{24}}  \left[2 s_{13}+ 2 s_{14}-s_{23}\right]\nonumber \\ 
&& + \frac{4 s_{14} s_{12}^2s_{23}}{ s_{34}^2  s_{134} s_{234} }  + \frac{s_{14}}{ s_{34}^2 s_{134}}  \Big[ +4  s_{12}s_{14} + 4 s_{14} s_{23}+ 4 s_{14}s_{24}\nonumber \\ 
&&+ 8 s_{12} s_{24}+ 4 s_{12}^2+ 4 s_{23} s_{24}+ 4 s_{24}^2\Big]\nonumber \\ 
&&- \frac{s_{14}}{ s_{34} s_{24} s_{134}}  \left[ 2 s_{12}s_{14} + s_{14} s_{23}- 2 s_{14}^2- s_{12} s_{23}- s_{12}^2\right]\nonumber \\ 
&&+ \frac{s_{14}}{ s_{34} s_{134}^2 } \left[ - 4 s_{12} s_{23} - 4 s_{12} s_{24}- 2 s_{12}^2 - 4 s_{23} s_{24}- 2 s_{23}^2- 2 s_{24}^2\right]\nonumber \\ 
&&- \frac{s_{14}}{ s_{234}} + \frac{s_{14}^2}{ s_{34}^2 s_{134}^2}  \left[- 4 s_{12} s_{23}- 4 s_{12} s_{24}- 2 s_{12}^2 - 4 s_{23} s_{24}- 2 s_{23}^2- 2 s_{24}^2\right]\nonumber \\ 
&&- \frac{1}{s_{12} s_{34} s_{234}}  \left[s_{13} s_{14}^2+ s_{13}^2 s_{14}- s_{13}^2 s_{23}+ s_{13}^3+ s_{14}^2 s_{23}+ s_{14}^3\right]\nonumber \\ 
&&- \frac{1}{s_{12} s_{34}}  \Big[s_{13} s_{14} - 3 s_{13} s_{23}- s_{13} s_{24}+ 2 s_{13}^2+ s_{14} s_{23}- s_{14} s_{24}\nonumber \\
&&+ 4 s_{14}^2+ s_{23} s_{24}+ s_{23}^2\Big]- \frac{1}{s_{12} s_{134}}  \Big[  s_{13} s_{24} - s_{14} s_{23}\nonumber \\ 
&&- 2 s_{14} s_{24}- s_{23} s_{24}+ s_{24}^2 \Big]- \frac{1}{s_{12}}  \left[s_{13}+ 2 s_{14} - s_{23}  - s_{24} \right]\nonumber \\ 
&&- \frac{s_{12}}{ s_{34} s_{134} s_{234}}  \left[  2 s_{14} s_{23} + 8 s_{14}^2 - 2 s_{12} s_{23} + 2 s_{12}^2 + 2 s_{23}^2\right]\nonumber \\ 
&&- \frac{s_{23}}{s_{34}^2  s_{234} } \Big[4 s_{13} s_{14} - 8 s_{12} s_{13} + 4 s_{13} s_{23} + 4 s_{13}^2 + 4 s_{14} s_{23} + 4 s_{12} s_{23}\nonumber \\ 
&& + 4 s_{12}^2\Big]+ \frac{s_{23}^2 }{s_{34}^2 s_{234}^2}  \left[ - 4 s_{13} s_{14} + 4 s_{12} s_{13} - 2 s_{13}^2 + 4s_{12} s_{14}  - 2 s_{14}^2 - 2 s_{12}^2\right]\nonumber \\ 
&&+ \frac{1}{s_{34}^2}  \Big[ 4 s_{13} s_{12}- 4 s_{13} s_{23}+ 4 s_{13} s_{24}- 2 s_{13}^2-4 s_{14} s_{12}  - 4 s_{14} s_{23}\nonumber \\ 
&& - 4 s_{14} s_{24} + 4 s_{12} s_{23}- 4 s_{12} s_{24}- 2 s_{12}^2- 2 s_{24}^2\Big]\nonumber \\ 
&&+ \frac{1}{s_{34} s_{134}}  \Big[  s_{12}s_{14} + 7 s_{14} s_{23}- s_{14} s_{24}+ 7 s_{14}^2+ s_{12} s_{23} + 7 s_{12} s_{24}\nonumber \\ 
&&+ 6 s_{12}^2+ 2 s_{23} s_{24} + s_{23}^2 + 3 s_{24}^2\Big]+ \frac{1}{s_{34} s_{234}}  \Big[-6  s_{12}s_{13}- 2 s_{13} s_{23} + 4 s_{13}^2\nonumber \\ 
&&  + 4 s_{14} s_{23}+ 6 s_{14}^2 + 5 s_{12}^2\Big]+ \frac{1}{s_{34}}  \left[3 s_{13}+ 2 s_{14}- 6 s_{12}- 6 s_{23} - s_{24}\right]\nonumber \\ 
&& + \frac{s_{23}}{ s_{24} s_{234}^2 } \left[ 2 s_{13} s_{14} - 2 s_{12}s_{13} + s_{13}^2- 2  s_{12}s_{14} + s_{14}^2 + s_{12}^2 \right]\nonumber \\
&&- \frac{1}{s_{24} s_{134} s_{234}}  \Big[  s_{12}s_{13} s_{14} - s_{13} s_{14} s_{23} - s_{13} s_{14}^2+  s_{12}s_{13} s_{23}+ 3  s_{12}^2s_{14}\nonumber \\ 
&&- 3  s_{12}s_{14}^2 + s_{14}^3- s_{12}^3 \Big]- \frac{1}{s_{24} s_{134}}  \left[   s_{12}s_{13}-s_{12} s_{14}  - s_{14}^2+ s_{12}^2 \right]\nonumber \\ 
&&- \frac{1}{s_{24} s_{234} } \Big[ 2 s_{12}s_{13} - 2 s_{13} s_{23} - s_{13}^2 - 3 s_{12}s_{14} - s_{14} s_{23}+ 4 s_{14}^2\nonumber \\ 
&&+ s_{12} s_{23} \Big]+ \frac{1}{s_{24} } \left[  2 s_{13}- 2 s_{14} + s_{34} \right]\nonumber \\ 
&&+ \frac{1}{s_{134}^2  }\left[- 2 s_{12} s_{23} - 2 s_{12} s_{24} - s_{12}^2 - 2 s_{23} s_{24} - s_{23}^2 - s_{24}^2\right]\nonumber \\ 
&&+ \frac{1}{s_{134} s_{234}}  \left[s_{13} s_{14} + 3s_{12} s_{13}  - s_{13}^2- 4 s_{12} s_{14} - s_{14}^2  - 3 s_{12}^2 \right]\nonumber \\ 
&&+ \frac{1}{s_{134} } \left[ - 3 s_{13}+ 6 s_{14} + 3 s_{23} - 2 s_{24}\right]+ 2\nonumber\\
&&+m_Qm_{\chi}\bigg[ -\frac{s_{12}}{s_{24} s_{134}}+\frac{3}{s_{34}}-\frac{s_{14}}{s_{34} s_{134}}+\frac{s_{14} s_{23}}{s_{24} s_{34}s_{134}}-\frac{2}{s_{134}}\nonumber\\
&& -\frac{s_{23}}{s_{24} s_{345}}+\frac{2 s_{23}}{s_{34} s_{345}}-\frac{1}{s_{345}}-\frac{4 s_{14}}{s_{12}s_{24}}+\frac{1}{s_{12}s_{34}}\left[s_{24}-s_{13}\right] \nonumber\\
&&+\frac{s_{14} s_{23}}{s_{12}s_{24} s_{34} }-\frac{s_{24}}{s_{12}s_{134}}-\frac{s_{14} }{s_{12}s_{34} s_{134} }\left[s_{23}+3s_{24} \right]\nonumber\\
&&-\frac{2s_{14}^2}{s_{12}s_{34}^2 s_{134} } \left[ s_{23} +s_{24} \right]+\frac{s_{13}}{s_{345}s_{12}}-\frac{s_{14} s_{23}}{s_{24} s_{345} s_{12}}\nonumber\\
&&-\frac{2 s_{13} s_{23}}{s_{34} s_{345} s_{12}}+\frac{2s_{23}^2}{s_{34}^2 s_{345} s_{12}} \left[s_{13} +s_{14}\right]\nonumber\\
&&+\frac{2}{ s_{12}s_{34}^2} \left[s_{13} s_{23}+s_{14} s_{23}-s_{13} s_{24}+s_{14} s_{24}\right]\bigg]\nonumber\\
&&+m_Q^2\bigg[\frac{s_{12}}{s_{24} s_{134}}-\frac{4}{s_{34}}+\frac{2}{s_{34} s_{134}} \left[2 s_{12}-s_{14}+s_{23}+3 s_{24}\right]+\frac{2 s_{23}}{s_{34}s_{345}}\nonumber\\
&&-\frac{1}{s_{134} s_{345}}\left[ s_{12}+s_{14}+s_{23}+s_{24}\right]+\frac{ s_{23}}{s_{24} s_{134} s_{345}}\left[s_{12}-s_{14}\right]\nonumber\\
&&-\frac{4 s_{23}}{s_{34} s_{134} s_{345}}\left[s_{23}+s_{14}\right]+\frac{1}{s_{345}}+\frac{1}{s_{34} s_{12}}\left[4 s_{14}-2 s_{23}-3 s_{24}\right]\nonumber\\
&&-\frac{2}{s_{34} s_{134}^2} \left[s_{12}^2+2  s_{12}s_{23}+2 s_{12}s_{24}+s_{23}^2+s_{24}^2+2 s_{23} s_{24}\right]\nonumber\\
&&+\frac{1}{s_{12} s_{134} }\left[2s_{23}-s_{24}\right]+\frac{1}{s_{12} s_{34} s_{134} }\left[2 s_{23}^2+5 s_{14} s_{23}+2 s_{24}^2-s_{14} s_{24}\right]\nonumber\\
&&+\frac{2}{s_{12} s_{34}^2 s_{134} } \left[s_{14}^2 s_{23}+ s_{14}^2 s_{24}\right]-\frac{2}{ s_{12}s_{34} s_{345}}\left[-2 s_{23}^2+s_{13} s_{23}-2 s_{14}s_{23}\right]\nonumber\\
&&-\frac{2 s_{23}^2}{ s_{12} s_{34}^2 s_{345}} \left[s_{13}+s_{14}\right]-\frac{2}{s_{12} s_{34}^2 }\left[ s_{13}s_{23}+s_{14} s_{23}-s_{13} s_{24}+s_{14} s_{24}\right]\nonumber\\
&&+\frac{2}{s_{12}}-\frac{2}{s_{12}^2}\left[s_{13}+s_{14}-s_{23}-s_{24}\right] -\frac{2 s_{23}}{s_{12} s_{345} }\nonumber\\
&&-\frac{2}{s_{12}^2 s_{34} } \left[s_{13}^2-2 s_{13}s_{23} +s_{14}^2+s_{23}^2+s_{24}^2-2 s_{14}s_{24}\right]\bigg]\nonumber\\
&&+m_Q^3m_{\chi}\bigg[ \frac{4}{s_{12}^2}-\frac{2}{s_{12} s_{34} s_{134}} \left[s_{23}+s_{24}\right]\bigg]+\frac{2m_Q^4}{s_{12} s_{34} s_{134}}\left[s_{23}+s_{24}\right]
\bigg\}\nonumber\\
&&+\order{\epsilon},\nonumber\\ \label{eq.E04ifm}
\eeqa
where $Q^2=-(p_1+p_3+p_4-p_2)^2$, $m_{\chi}=\sqrt{Q^2}$, $s_{134}=s_{13}+s_{14}+s_{34}$ and $s_{234}=s_{34}-s_{23}-s_{24}$.

In order to account for the triple collinear limits that involve a massless $q\bar{q}$ pair and an initial state gluon in those sub-leading colour amplitudes in which the gluon is placed between the quark and the antiquark in the gluon chain, we employ the following $\wt{E}_4^0$ antenna
\beqa
\wt{E}_4^0(\Q{1},\q{3},\qb{4},\gli{2})&=&\frac{1}{\left(Q^2+m_Q^2\right)^2}\bigg\{ \frac{1}{s_{23} s_{24}}\Big[ -2 s_{12} s_{13}-2 s_{12} s_{14}-2 s_{12} s_{34}+2 s_{12}^2\nonumber\\
&&+2 s_{13} s_{34}+2 s_{13}^2+2 s_{14} s_{34}+2 s_{14}^2 \Big]+\frac{s_{23}}{s_{234}^2 s_{24}}\Big[-2 s_{12} s_{13}-2 s_{12} s_{14}\nonumber\\
&&+s_{12}^2+2 s_{13} s_{14}+s_{13}^2+s_{14}^2\Big]+\frac{s_{24}}{s_{23} s_{234}^2} \Big[-2 s_{12} s_{13}-2 s_{12} s_{14}+s_{12}^2\nonumber\\
&&+2 s_{13} s_{14}+s_{13}^2+s_{14}^2\Big]-\frac{1}{s_{234} s_{24}}\Big[4 s_{12} s_{13}+2 s_{12} s_{14}+s_{12} s_{23}-2 s_{12}^2\nonumber\\
&&-2 s_{13} s_{14}-s_{13} s_{23}-2s_{13}^2-s_{14} s_{23}\Big]-\frac{1}{s_{23} s_{234}}\Big[2 s_{12} s_{13}+4 s_{12} s_{14}\nonumber\\
&&+s_{12} s_{24}-2s_{12}^2-2 s_{13} s_{14}-s_{13} s_{24}-s_{14} s_{24}-2 s_{14}^2\Big]\nonumber\\   
&&-m_Q m_{\chi}\left[ \frac{2 s_{24}}{s_{23} s_{234}}+\frac{4 s_{234}}{s_{23} s_{24}}+\frac{2 s_{23}}{s_{234} s_{24}}+\frac{4}{s_{23}}+\frac{4}{s_{24}}\right] \bigg\}+\order{\epsilon},\nonumber\\ \label{eq.E04tifm}
\eeqa
with $Q^2$, $m_{\chi}$, $s_{134}$ and $s_{234}$ given as above. Both E-type antennae are normalised to the tree-level two-parton matrix element
\beq
\left| \cm_2(\tilde{\chi} g  \rightarrow \tilde{g})\right|^2=4 (1-\epsilon)\left[Q^2+m_Q^2\right]^2.
\eeq

\section{Infrared limits of massive NNLO antennae}\label{sec.limits}
The factorisation properties of QCD tree-level squared amplitudes have been extensively studied in~\cite{Campbell:1997hg,Catani:1998nv,Catani:1999ss,deFlorian:2001zd}. While at NLO only single soft and collinear singularities may arise, at NNLO, several double unresolved configurations involving two soft and/or collinear particles can arise. In the first part of this section, we shall list all double unresolved factors arising in the unresolved limits of the antennae needed to construct our subtraction term presented in section \ref{sec.subterms}, while in the second part we shall list the infrared limits of all required four-parton antennae. The single unresolved factors have been already discussed in \cite{Daleo:2006xa,Abelof:2011jv,Abelof:2011ap} and will be given in the appendix \ref{sec.singlefactors} for completeness.

\subsection{Double unresolved factors}
In general, when two particles are unresolved in a tree-level process,  a variety of different configurations can arise:
\begin{itemize}
\item[{(1)}]  two soft particles,
\item[{(2)}]  two pairs of collinear particles,
\item[{(3)}]  three collinear particles,
\item[{(4)}]  one soft and two collinear.
\end{itemize}

As we saw in section \ref{sec.formalism}, the resulting double unresolved configurations of the colour-ordered matrix element squared need to be separated into three categories depending on the colour connections of the unresolved partons and the hard radiators associated with them. In the following, we shall describe only the double unresolved factors encountered in the context of this paper, which correspond to the configurations (1), (2) and (3) listed above, but not to (4), since the process $gg\rightarrow Q\bar{Q}q\bar{q}$ does not have any soft-and-collinear limits.

For unresolved particles that are disjoint in the colour chain, which arise in the item $(d)$ as presented in section \ref{sec.formalism}, the colour-ordered matrix elements squared factorise into the product of two disjoint single unresolved factors multiplied by a reduced matrix element with two partons less than the original colour-ordered matrix element squared. In the process that we are considering in this paper, the only double unresolved colour-unconnected limits that can occur are double collinear (anti) quark-gluon limits, with each single collinear pair given by a final state (anti) quark and an initial state gluon. The unresolved factor associated to this limit is a product of two splitting functions of the type given in eqs.(\ref{eq.splitting1}-\ref{eq.splitting7}). Note that none of the four-parton antennae required for our subtraction term captures these double collinear singularities. These are accounted for by subtraction terms involving the product of two three-parton antennae.

The colour-connected double unresolved limits the partonic process $gg\to Q\bar{Q}q\bar{q}$ develops are:
\begin{enumerate}
\item [A)] Triple collinear limits involving an initial state gluon and a massless final state quark-antiquark pair.
\item [B)] Double soft limits of a massless final state $q\bar{q}$ pair emitted between massive or massless radiators. 
\end{enumerate}
These colour-connected double unresolved limits are captured by four-parton antenna functions and arise in subtraction terms ${\rm d}\sigma_{NNLO}^{S,b}$. In the following, we shall give the massive and massless double unresolved factors associated with these two types of limits.\\

\parindent 0em

A) {\bf Massless triple collinear factors}\\
In those regions of phase space where three colour-connected massless partons $(a,b,c)$ become collinear, a generic colour-ordered amplitude squared denoted by $|\cm_n(\ldots,a,b,c,\ldots)|^2$ factorises as:
\beq
|\cm_{n}(\ldots,a,b,c,\ldots)|^2 \rightarrow P_{abc \rightarrow P}|\cm_{n-2}(\ldots,P,\ldots)|^2.
\eeq
where the three colour-connected final state particles $(a,b,c)$ cluster to form a single parent particle $P$. The triple collinear splitting function for partons $a$, $b$ and $c$ clustering to form the parent parton $P$ is generically denoted by,
\beq
P_{abc \rightarrow P}(w,x,y,s_{ab},s_{ac},s_{bc},s_{abc}),
\eeq
where $w$, $x$ and $y$ are the momentum fractions of the clustered partons,
\beq\label{momfrac}
p_a=wp_P, \qquad p_b=xp_P, \qquad p_c=yp_P, \hspace{10mm} \mbox{with }\hspace{2mm} w+x+y=1.
\eeq
In addition to its dependence on the momentum fractions carried by the clustering partons, the splitting function also depends on the invariant masses of parton-parton pairs and the invariant mass of the whole cluster. The explicit forms of the triple collinear splitting functions  $P_{abc \rightarrow P}$ are obtained by retaining terms in the colour-ordered matrix element squared that possess two of the `small' denominators $s_{ab}$, $s_{ac}$, $s_{bc}$ and $s_{abc}$.

\parindent 1.5em

The triple collinear limits which have to be considered here are those involving the massless final state quark-antiquark pair and one of the initial state gluons. The splitting functions for these types of (initial-final) triple collinear limits can be obtained from the analoguous limit where the three collinear particles are in the final state~\cite{Campbell:1997hg,deFlorian:2001zd}.

The clustering of a gluon with a quark-antiquark pair into a parent gluon has two distinct functions depending the colour connection of the collinear particles. In leading colour contributions where the gluon is emitted ``outside'' the quark-antiquark pair, the following non-abelian splitting function is obtained
\beqa
\lefteqn{P_{g{\bar q}q \rightarrow G}(w,x,y,s_{g\bar q},s_{\bar q q},s_{g{\bar q}q}) =}\nonumber\\
&-& \frac{1}{s_{g{\bar q}q}^2} \left( 4\frac{s_{g{\bar q}}}{s_{\bar q q}}+(1-\e)\frac{s_{\bar q q}}{s_{g{\bar q}}} + (3-\e) \right) - \frac{2 \left( xs_{g{\bar q}q}-(1-w)s_{g{\bar q}} \right)^2}{s_{\bar q q}^2s_{g{\bar q}q}^2(1-w)^2} \nonumber \\
&+& \frac{1}{s_{g{\bar q}}s_{g{\bar q}q}} \left( \frac{(1-y)}{w(1-w)}-y-2w-\e -\frac{2x(1-y)(y-w)}{(1-\e)w(1-w)} \right) \nonumber \\
&+&  \frac{1}{s_{g{\bar q}}s_{\bar q q}} \left( \frac{x\left( (1-w)^3-w^3\right)} {w(1-w)} -\frac{2x^2 \left( 1-yw-(1-y)(1-w) \right)}{(1-\e)w(1-w)} \right)\nonumber \\
&+& \frac{1}{s_{{\bar q}q}s_{g{\bar q}q}} \left( \frac{(1+w^3+4xw)}{w(1-w)}+\frac{2x \left( w(x-y)-y(1+w) \right)}{(1-\e)w(1-w)} \right).\label{eq.triplecoll}
\eeqa
In those sub-leading colour pieces where the gluon is emitted ``between'' the quark-antiquark pair in the colour chain, one obtains a QED-like splitting function 
\begin{eqnarray}
\lefteqn{\tilde{P}_{q g {\bar q} \rightarrow G}(w,x,y,s_{qg},s_{g\bar q},s_{\bar q q},s_{qg{\bar q}}) =}\nonumber\\
&-& \frac{1}{s_{qg{\bar q}}^2} \left( (1-\e)\frac{s_{q{\bar q}}}{s_{qg}} +1 \right)+ \frac{1}{s_{g\bar q}s_{qg}} \left( (1+x^2)-\frac{x+2wy}{1-\e} \right)\nonumber \\
&-& \frac{1}{s_{qg}s_{qg{\bar q}}} \left( 1+2x+\e-\frac{2(1-y)}{(1-\e)} \right)+ ( s_{qg} \leftrightarrow s_{g\bar q}, w \leftrightarrow y).\label{eq.triplecollab}
\end{eqnarray}

The triple collinear splitting functions given in eqs.(\ref{eq.triplecoll}) and (\ref{eq.triplecollab}) correspond to configurations in which all three collinear particles are outgoing. As mentioned above, in the present calculation we do not deal with these limits (indeed, there are not enough massless particles in the final state to have final-final triple collinear limits). Instead, we deal with triple collinear limits where one of the partons (a gluon) is in the initial state. In this case, initial-final triple collinear splitting functions arise. They are related  to their final-final counterparts as follows~\cite{deFlorian:2001zd}  
\beqa
&&\hspace{-5mm}P_{g\bar{q}q\leftarrow G}(z_3,z_2,z_1,s_{\hat{g}\bar{q}},s_{q\bar{q}},s_{\hat{g}\bar{q}q})=\nonumber\\
&&\hspace{30mm}P_{g\bar{q}q\rightarrow G}(1/z_3,-z_2/z_3,-z_1/z_3,-s_{g\bar{q}},s_{q\bar{q}},s_{q\bar{q}}-s_{g\bar{q}}-s_{qg})\nonumber\\
&&\hspace{-5mm}\tilde{P}_{qg\bar{q}\leftarrow G}(z_1,z_3,z_2,s_{q\hat{g}},s_{\hat{g}\bar{q}},s_{q\bar{q}},s_{q\hat{g}\bar{q}})=\nonumber\\
&&\hspace{30mm}\tilde{P}_{qg\bar{q}\rightarrow G}(-z_1/z_3,1/z_3,-z_2/z_3,-s_{qg},-s_{g\bar{q}},s_{q\bar{q}},s_{q\bar{q}}-s_{g\bar{q}}-s_{qg}),\nonumber\\
\label{eq.triplecollif}
\eeqa
where $z_1$ and $z_2$ are the momentum fractions final state quark and antiquark respectively, and $z_3=1-z_1-z_2$.\\
\parindent 0em

B) {\bf Massive double soft factor}\\
When a massless quark-antiquark pair becomes soft between two hard radiators, sub-amplitudes squared factorise into a double soft factor and a reduced matrix element squared with the quark-antiquark pair removed from it. This double soft factor also depends on the masses of the hard radiators. If the soft quark-antiquark pair is denoted by $(c,d)$ and the hard radiators $(a,b)$ have masses $m_{a}$ and $m_{b}$, the double soft factor is given by, 
\beqa
\soft{a}{c}{d}{b}(m_a,m_b)&=&\frac{2(s_{ab}s_{cd}-s_{ac}s_{bd}-s_{bc}s_{ad})}{s_{cd}^2(s_{ac}+s_{ad})(s_{bc}+s_{bd})}+\frac{2}{s_{cd}^2}\left[ \frac{s_{ac}s_{ad}}{(s_{ac}+s_{ad})^2} +  \frac{s_{bc}s_{bd}}{(s_{bc}+s_{bd})^2}\right]\nonumber\\
&&-\frac{2m_{a}^2}{s_{cd}(s_{ac}+s_{ad})^2}-\frac{2m_{b}^2}{s_{cd}(s_{bc}+s_{bd})^2}.\label{eq.seik}
\eeqa
This factor is obtained by setting $p_c \to \lambda p_c,\:\:p_{d} \to \lambda p_d$ with $\lambda \to 0$ in the matrix elements squared and was derived in~\cite{Bernreuther:2011jt} as the soft $q\bar{q}$ limit of the massive final-final antenna $B_4^0(Q,\bar{q},q,\bar{Q})$ and in~\cite{Abelof:2011ap} using current algebra. Note also that, the corresponding massless factor \cite{Campbell:1997hg,GehrmannDeRidder:2005cm} can be obtained from this massive one by setting the masses to zero.

\parindent 1.5em

\subsection {Single and double unresolved limits of the four-parton antennae}\label{subsec.limitsA}
In their infrared limits, the four-parton antennae of B, E and G types required to compute the double real contributions to heavy quark pair production in hadronic collisions due to the partonic process $gg\rightarrow Q\bar{Q}q\bar{q}$, and defined in section \ref{sec.antennae}, yield the universal single and double unresolved factors defined above. Additionally, some of these four-parton antennae yield angular correlation terms in single collinear limits. These angular terms (denoted by $ang.$ below) arise when a gluon splits into a quark-antiquark pair or into two gluons, and the way in which they are dealt with in the antenna subtraction method has been explained in detail in ~\cite{Abelof:2011ap,GehrmannDeRidder:2007jk,Glover:2010im,Weinzierl:2006ij} and will not be recalled here.

In addition to the conventions already defined in section \ref{sec.antennae} for the labelling of the partons present in antenna function, we shall denoted with $(ij)_{\alpha}$ the momentum of the parent partons in collinear limits. Thus, partons labelled with $(ij)_{\alpha}$ will have momentum $p_i+p_j$ in final-final collinear limits, whereas their momentum will be given by $p_i-p_j$ in the initial-final case. The corresponding splitting functions are denoted as $P_{ij\rightarrow(ij)}(z)$ when both collinear particles are in the final state, and with $P_{ij\leftarrow(ij)}(z)$ in initial-final collinear limits. They are listed in appendix \ref{sec.singlefactors}.\\

\parindent 0em

{\bf{Massive final-final B-type antennae}}\\
The massive final-final B-type antenna given in eq.(\ref{eq.B04ffm}) is only singular when the massless $q\bar{q}$ pair is soft or collinear. The behaviour of this antenna in each of these limits is given by
\beqa
&&B_4^0(\Q{1},\qb{4},\q{3},\Qb{2})\stackrel{^{\q{3},\qb{4}\rightarrow0}}{\longrightarrow}\soft{1}{3}{4}{2}(m_Q,m_Q)\label{eq.softfactb04}\\
&&B_4^0(\Q{1},\qb{4},\q{3},\Qb{2})\stackrel{^{\q{3}||\qb{4}}}{\longrightarrow}\frac{1}{s_{34}}P_{q\bar{q}\rightarrow G}(z)A_3^0(\Q{1},\gl{(34)},\Qb{2})+{\text{ang.}}.
\eeqa

{\bf{Massless initial-initial G-type antennae}}\\
The massless initial-initial G-type antenna given in the appendix in eq.(\ref{eq.G04ii}) has the following infrared limits 
\beqa
&&G_4^0(\gli{1},\q{3},\qb{4},\gli{2})\stackrel{^{\q{3},\qb{4}\rightarrow0}}{\longrightarrow}\soft{1}{3}{4}{2}(0,0)\label{eq.softfactg04}\\
&&G_4^0(\gli{1},\q{3},\qb{4},\gli{2}) \stackrel{^{\gli{1}||\q{3}||\qb{4}}}{\longrightarrow}P_{g\bar{q}q\leftarrow G}(z_3,z_2,z_1,s_{13},s_{34},s_{134})\\
&&G_4^0(\gli{1},\q{3},\qb{4},\gli{2}) \stackrel{^{\gli{2}||\q{3}||\qb{4}}}{\longrightarrow}P_{g\bar{q}q\leftarrow G}(z_3,z_2,z_1,s_{24},s_{34},s_{234})\\
&&G_4^0(\gli{1},\q{3},\qb{4},\gli{2})\stackrel{^{\q{3}||\qb{4}}}{\longrightarrow}\frac{1}{s_{34}}P_{q\bar{q}\rightarrow G}(z)F_3^0(\gli{1},\gl{(34)},\gli{2})+{\text{ang.}}\\
&&G_4^0(\gli{1},\q{3},\qb{4},\gli{2})\stackrel{^{\gli{1}||\q{3}}}{\longrightarrow}\frac{1}{s_{13}}P_{q\bar{q}\leftarrow G}(z)G_3^0(\gli{2},\qb{4},\widehat{(13)}_{\bar{q}})\\
&&G_4^0(\gli{1},\q{3},\qb{4},\gli{2})\stackrel{^{\gli{2}||\qb{4}}}{\longrightarrow}\frac{1}{s_{24}}P_{q\bar{q}\leftarrow G}(z)G_3^0(\gli{1},\q{3},\widehat{(24)}_{q}).
\eeqa

{\bf{Massive intial-final E and $\tilde{E}$ -type antennae}}\\
The initial-final E-type antenna given in eq.(\ref{eq.E04ifm}) has a soft $q\bar{q}$ limit, a triple collinear limit, a final-final and an initial-final single collinear limits. The behaviour of this antenna in each of these limits is
\beqa
&&E_4^0(\Q{1},\q{3},\qb{4},\gli{2})\stackrel{^{\q{3},\qb{4}\rightarrow0}}{\longrightarrow}\soft{1}{3}{4}{2}(m_Q,0)\label{eq.softfacte04}\\
&&E_4^0(\Q{1},\q{3},\qb{4},\gli{2})\stackrel{^{\gli{2}||\q{3}||\qb{4}}}{\longrightarrow}P_{g\bar{q}q\leftarrow G}(z_3,z_2,z_1,s_{24},s_{34},s_{234})\\
&&E_4^0(\Q{1},\q{3},\qb{4},\gli{2})\stackrel{^{\q{3}||\qb{4}}}{\longrightarrow}\frac{1}{s_{34}}P_{q\bar{q}\rightarrow G}(z)D_3^0(\Q{1},\gl{(34)},\gli{2})+{\text{ang.}}\label{eq.e04coll}\\
&&E_4^0(\Q{1},\q{3},\qb{4},\gli{2})\stackrel{^{\gli{2}||\qb{4}}}{\longrightarrow}\frac{1}{s_{24}}P_{q\bar{q}\leftarrow G}(z)E_3^0(\Q{1},\q{3},\widehat{(24)}).
\eeqa
The initial-final massive antenna $D_3^0(Q,g,\hat{g})$ is obtained by crossing to the initial state one of the gluons in its final-final counterpart. As it can be seen in eq.(\ref{eq.e04coll}), it is the antenna onto which $E_4^0(Q,q,\bar{q},\hat{g})$ collapses to in its single $q||\bar{q}$ collinear limit. It will be given in the appendix \ref{sec.3partonantennae}.

\parindent 1.5em

The sub-leading colour antenna $\wt{E}_4^0$ has a QED-like triple collinear and two initial-final single collinear limits:
\beqa
&&\wt{E}_4^0(\Q{1},\q{3},\qb{4},\gli{2})\stackrel{^{\gli{2}||\q{3}||\qb{4}}}{\longrightarrow}\tilde{P}_{qg\bar{q}\leftarrow G}(z_1,z_3,z_2,s_{23},s_{24},s_{34},s_{234}),\\
&&\wt{E}_4^0(\Q{1},\q{3},\qb{4},\gli{2})\stackrel{^{\gli{2}||\qb{4}}}{\longrightarrow}\frac{1}{s_{24}}P_{q\bar{q}\leftarrow G}(z)E_3^0(\Q{1},\q{3},\widehat{(24)}_{q})\\
&&\wt{E}_4^0(\Q{1},\q{3},\qb{4},\gli{2})\stackrel{^{\gli{2}||\qb{3}}}{\longrightarrow}\frac{1}{s_{23}}P_{q\bar{q}\leftarrow G}(z)E_3^0(\Q{1},\qb{4},\widehat{(23)}_{\bar{q}}).
\eeqa

\section{Top quark pair production at the LHC}\label{sec.subterms}
In this section we shall present the double real emission contributions to $t\bar{t}$ production at the LHC due to the process $gg\rightarrow Q\bar{Q}q\bar{q}$. Together with these, we shall give their corresponding antenna subtraction terms, which capture all single and double unresolved limits of the leading and subleading colour pieces of the real radiation matrix elements squared.

\subsection{Conventions}
To facilitate the reading of our expressions, we shall closely follow the notation in~\cite{Abelof:2011ap,Abelof:2011jv} for matrix elements and subtractions terms. The main points of our conventions recalled here for clarity reasons, are the following: The matrix elements denoted with ${\cal M}$ represent colour-ordered sub-amplitudes in which the coupling constants and colour factors are omitted. Furthermore, to explicitly visualise the colour connection between particles in these colour-ordered amplitudes, a double semicolon is used in the labeling of the partons present in a given matrix element. This double semicolon is used for separating chains of colour-connected partons. Partons within a pair of double semicolons belong to a same colour chain, and adjacent partons within a colour chain are colour-connected. An antiquark (or an initial state quark) at the end of a colour chain and a like flavour quark (or initial state antiquark) at the beginning of a different colour chain are also colour-connected since the two chains merge in the collinear limit where the $q\bar{q}$ clusters into a gluon. We also denote gluons which are photon-like and only couple to quark lines with the index $\gamma$ instead of $g$. In sub-amplitudes where all gluons are photon-like no semicolons are used, since the concept of colour connection is not meaningful. A hat over the label of a certain parton indicates that it is an initial state particle (for example, $\qi{1}$ is an initial state quark with momentum $p_1$).

Concerning the notation in the subtraction terms, the conventions for the reduced matrix elements are the same as those for the real radiation matrix elements discussed above. The remapped final-state momenta are denoted with tildes and the remapped momenta of initial state hard radiators are denoted by a bar and a hat, as used in other papers~\cite{Abelof:2011ap,GehrmannDeRidder:2011aa,Glover:2010im}. In the four-parton antenna functions the hard radiators are ``on the edges'' and the uresolved particles are ``in the middle''. \\

\subsection{Double real radiation contributions}
For two incoming hadrons, $H_1,H_2$, the hadronic heavy quark pair production cross section may be written as 
\beq\label{eq.hadroncross}
{\rm d} \sigma= \sum_{a,b} \int \frac{{\rm d} \xi_1}{\xi_1} \frac{{\rm d} \xi_2}{\xi_2}\, f_{a/1}(\xi_1,\mu_F) \,f_{b/2}(\xi_2,\mu_F)\, {\rm d} \hat{\sigma}_{ab}(\xi_1H_1,\xi_2H_2,\mu_F,\mu_R)\ .
\eeq
$\xi_1$ and $\xi_2$ are the momentum fractions of the partons of species $a$ and $b$ in both incoming hadrons, $f_i$ being the corresponding parton distribution functions. ${\rm d} \hat{\sigma}_{ab}$ denotes the parton-level scattering cross section for incoming partons $a$ and $b$ which depends on the the renormalisation and factorisation scales denoted by $\mu_R$ and $\mu_{F}$ respectively. The partonic cross section ${\rm d }\hat{\sigma}_{ab}$ has a perturbative expansion in the strong coupling $\alpha_{s}$ which itself depends on the renormalisation scale $\mu_{R}$.

Following the general factorisation formula given in eq.(\ref{eq.hadroncross}) (if we omit the renormalisation and factorisation scale dependences) the contribution of the partonic process $gg\rightarrow Q\bar{Q}$ to the leading order cross section for heavy $Q\bar{Q}$ production in hadronic collisions is
\beq
{\rm d}\sigma_{gg\rightarrow Q\bar{Q}}^{LO}=\int \frac{{\rm d}\xi_1}{\xi_1}\frac{{\rm d}\xi_2}{\xi_2}  f_g(\xi_1) f_{g}(\xi_2)\ds^{LO}_{gg\rightarrow Q\bar{Q}}.
\eeq

The leading order partonic differential cross section, written in terms of colour-ordered matrix elements, is given by
\beqa
&&\ds^{LO}_{gg\rightarrow Q\bar{Q}}=\norm_{LO}\dphi_2(p_1,p_2;p_3,p_4)\bigg[ N_c \left( |\cm_4(\Q{1},\gli{3},\gli{4},\Qb{2})|^2 + |\cm_4(\Q{1},\gli{4},\gli{3},\Qb{2})|^2 \right)\nonumber\\ &&\hspace{1.5in}-\frac{1}{N_c}|\cm_4(\Q{1},\phin{3},\phin{4},\Qb{2})|^2\bigg] J_2^{(2)}(p_1,p_2).
\eeqa
In this equation, the normalisation factor $\norm_{LO}$ reads
\beq\label{eq.norm}
\norm_{LO}=\frac{1}{2s}\times \left( \frac{\alpha_s}{2\pi} \right)^2  \frac{\bar{C}(\epsilon)^2}{C(\epsilon)^2}\times (N_c^2-1)\times \frac{1}{4(N_c^2-1)^2},
\eeq
where $s$ is the hadronic center of mass energy, $(N_c^2-1)$ comes from the colour sum, the factor $1/4(N_c^2-1)^2$ accounts for the averaging over the spin and colour of the incoming two gluons, and $(1/2s)$ is the hadron-hadron flux factor. The coupling is defined as usual: $\alpha_s=g^2/4\pi$, $C(\epsilon)=(4\pi)^{\e}e^{-\e\gamma}/8\pi^2$, and $\bar{C}(\epsilon)=(4\pi)^{\epsilon}e^{-\epsilon \gamma}$. This way of expressing the coupling factors, with each power of $\alpha_s$ accompanied by a power of $\bar{C}(\epsilon)$, keeps the coupling dimensionless in dimensional regularisation. 
The phase space for the production of a $Q \bar{Q}$ pair of momenta $p_1$ and $p_2$ is given by $\dphi_2(p_1,p_2;p_3,p_4)$. In it, $p_3$ and $p_4$ are the momenta of the initial state gluons. 
The jet function denoted by $J_{2}^{(2)}$ ensures that the top and the antitop are in two separate jets.

The colour-ordered amplitudes $\cm_4$ are related to the full amplitude $M_4^0(\Q{1},\Qb{2},\gli{3},\gli{4})$ through the colour decomposition
\beqa
M_4^0(\Q{1},\Qb{2},\gli{3},\gli{4})&=&g^2(\sqrt{2})^2\bigg( \left(T^{a_3}T^{a_4}\right)_{i_1i_2}\cm_4(\Q{1},\gli{3},\gli{4},\Qb{2})\nonumber\\
&&+\left(T^{a_4}T^{a_3}\right)_{i_1i_2}\cm_4(\Q{1},\gli{4},\gli{3},\Qb{2})\bigg),\label{eq.cdec4}
\eeqa
and the colour-ordered amplitude with photon-like gluons is given by
\beq\label{eq.cdec42}
\cm_4(\Q{1},\phin{3},\phin{4},\Qb{2})=\cm_4(\Q{1},\gli{3},\gli{4},\Qb{2})+\cm_4(\Q{1},\gli{4},\gli{3},\Qb{2}).
\eeq

The contribution of the $gg\rightarrow Q\bar{Q}q\bar{q}$ partonic channel to the double real radiation cross section for heavy quark pair production in hadronic collisions is
\beq\label{eq.hadronic}
{\rm d}\sigma^{RR}_{gg\rightarrow Q\bar{Q}q\bar{q}}=\int \frac{{\rm d}\xi_1}{\xi_1}\frac{{\rm d}\xi_2}{\xi_2} f_g(\xi_1)f_g(\xi_2)\ds^{RR}_{gg\rightarrow Q\bar{Q}q\bar{q}}
\eeq
with the partonic cross section given by
\beq\label{eq.partonic1}
\ds^{RR}_{gg\rightarrow Q\bar{Q}q\bar{q}}={\cal N}_{NNLO}^{RR}{\rm d}\Phi_4(p_1,p_2,p_3,p_4;p_5,p_6)|M^0_6(\Q{1},\Qb{2},\q{3},\qb{4},\gli{5},\gli{6})|^2 J_2^{(4)}(p_1,p_2,p_3,p_4).
\eeq
The full matrix element squared $|M^0_6(\ldots)|^2$ is summed over spin and colour. The normalisation factor ${\cal N}_{NNLO}^{RR}$ accounts for the spin and colour averaging, the flux factor and the sum over the possible flavours of the massless $q\bar{q}$ pair. The jet function $J_{2}^{(4)}$ corresponds to the selection criteria of a $2$-jet event: out of four partons, from which two are a $Q \bar{Q}$ pair, an event with two jets is built. Each of these two jets has the heavy quark $Q$ or the heavy antiquark $\bar{Q}$ in it. 
The additional partons present in a jet are either theoretically unresolved (soft or collinear) or not ``seen'' by the experimental resolution criteria (i.e not identified as a separate jet).

The full matrix element for the process $gg\rightarrow Q\bar{Q}q\bar{q}$ has the following colour decomposition~\cite{Abelof:2011jv}
\beqa
\lefteqn{M^0_6(\Q{1},\Qb{2},\q{3},\qb{4},\gli{5},\gli{6})=}\nonumber\\
&& g^4(\sqrt{2})^2\sum_{(i,j)\in P(5,6)}\bigg[(T^{a_i}T^{a_j})_{i_1i_4}\delta_{i_3,i_2} \cm_6(\Q{1},\gli{i},\gli{j},\qb{4};;\q{3},\Qb{2})\nonumber\\
&&\hspace{1.2in} +(T^{a_i})_{i_1i_4}(T^{a_j})_{i_3 i_2} \cm_6(\Q{1},\gli{i},\qb{4};;\q{3},\gli{j},\Qb{2})\nonumber\\
&&\hspace{1.2in}+\delta_{i_1,i_4}(T^{a_i}T^{a_j})_{i_3 i_2} \cm_6(\Q{1},\qb{4};;\q{3},\gli{i},\gli{j},\Qb{2})\nonumber\\
&&\hspace{1.2in}-\frac{1}{N_c}(T^{a_i}T^{a_j})_{i_1i_2}\delta_{i_3,i_4} \cm_6(\Q{1},\gli{i},\gli{j},\Qb{2};;\q{3},\qb{4})\nonumber\\
&&\hspace{1.2in}-\frac{1}{N_c}(T^{a_i})_{i_1i_2}(T^{a_j})_{i_3 i_4} \cm_6(\Q{1},\gli{i},\Qb{2};;\q{3},\gli{j},\qb{4})\nonumber\\
&&\hspace{1.2in}-\frac{1}{N_c}\delta_{i_1,i_2}(T^{a_i}T^{a_j})_{i_3i_4} \cm_6(\Q{1},\Qb{2};;\q{3},\gli{i},\gli{j},\qb{4})\bigg].\label{eq.colourdec}
\eeqa
Squaring eq.(\ref{eq.colourdec}), summing and averaging over spin, colour, and quark flavour, and plugging our result in eq.(\ref{eq.partonic1}) allows us to write the partonic double real radiation cross section in the following form
\beqa
\lefteqn{\ds_{gg\rightarrow Q\bar{Q}q\bar{q}}=\norm_{LO} N_F \left( \frac{\alpha_s}{2\pi} \right)^2\frac{\bar{C}(\epsilon)^2}{C(\epsilon)^2} \dphi_4(p_1,p_2,p_3,p_4;p_5,p_6) \sum_{(i,j)\in P(5,6)}\bigg[} \nonumber\\
&&\hspace{0.2in} N_c^2 \Big( |\cm_6(\Q{1},\gli{i},\gli{j},\qb{4};;\q{3},\Qb{2})|^2+ | \cm_6(\Q{1},\gli{i},\qb{4};;\q{3},\gli{j},\Qb{2})|^2\nonumber\\
&&\hspace{0.5in}\vspace{-1in}+|\cm_6(\Q{1},\qb{4};;\q{3},\gli{i},\gli{j},\Qb{2})|^2\Big)\nonumber\\
&&\hspace{0.2in}+| \cm_6(\Q{1},\gli{i},\gli{j},\Qb{2};;\q{3},\qb{4})|^2+|\cm_6(\Q{1},\gli{i},\Qb{2};;\q{3},\gli{j},\qb{4})|^2\nonumber\\
&&\hspace{0.2in}+|\cm_6(\Q{1},\Qb{2};;\q{3},\gli{i},\gli{j},\qb{4})|^2-|\cm_6(\Q{1},\gli{i},\gli{j},\qb{4};;\q{3},\Qb{2})|^2\nonumber\\
&&\hspace{0.2in}-| \cm_6(\Q{1},\gli{i},\qb{4};;\q{3},\gli{j},\Qb{2})|^2-| \cm_6(\Q{1},\qb{4};;\q{3},\gli{i},\gli{j},\Qb{2})|^2\nonumber\\
&&\hspace{0.2in} +2\re(\cm_6(\Q{1},\gli{i},\gli{j},\qb{4};;\q{3},\Qb{2})\cm_6(\Q{1},\qb{4};;\q{3},\gli{i},\gli{j},\Qb{2})^{\dagger})\nonumber\\
&&\hspace{0.2in} +2\re(\cm_6(\Q{1},\gli{i},\gli{j},\qb{4};;\q{3},\Qb{2})\cm_6(\Q{1},\qb{4};;\q{3},\gli{j},\gli{i},\Qb{2})^{\dagger})\nonumber\\
&&\hspace{0.2in} +\re(\cm_6(\Q{1},\gli{i},\qb{4};;\q{3},\gli{j},\Qb{2}) \cm_6(\Q{1},\gli{j},\qb{4};;\q{3},\gli{i},\Qb{2})^{\dagger})\nonumber\\
&&\hspace{0.2in} -\re(\cm_6(\Q{1},\gli{i},\gli{j},\qb{4};;\q{3},\Qb{2})\cm_6(\Q{1},\gli{j},\gli{i},\qb{4};;\q{3},\Qb{2})^{\dagger})\nonumber\\
&&\hspace{0.2in} -\re( \cm_6(\Q{1},\qb{4};;\q{3},\gli{i},\gli{j},\Qb{2})\cm_6(\Q{1},\qb{4};;\q{3},\gli{j},\gli{i},\Qb{2})^{\dagger})\nonumber\\
&&\hspace{0.2in} -2\re(\cm_6(\Q{1},\gli{i},\gli{j},\qb{4};;\q{3},\Qb{2})\cm_6(\Q{1},\gli{i},\gli{j},\Qb{2};;\q{3},\qb{4})^{\dagger})\nonumber\\
&&\hspace{0.2in} -2\re(\cm_6(\Q{1},\gli{i},\qb{4};;\q{3},\gli{j},\Qb{2})\cm_6(\Q{1},\gli{i},\gli{j},\Qb{2};;\q{3},\qb{4})^{\dagger})\nonumber\\
&&\hspace{0.2in} -2\re( \cm_6(\Q{1},\qb{4};;\q{3},\gli{i},\gli{j},\Qb{2})\cm_6(\Q{1},\gli{i},\gli{j},\Qb{2};;\q{3},\qb{4})^{\dagger})\nonumber\\
&&\hspace{0.2in} -2\re( \cm_6(\Q{1},\gli{i},\gli{j},\qb{4};;\q{3},\Qb{2})\cm_6(\Q{1},\gli{i},\Qb{2};;\q{3},\gli{j},\qb{4})^{\dagger})\nonumber\\
&&\hspace{0.2in} -2\re(\cm_6(\Q{1},\gli{i},\qb{4};;\q{3},\gli{j},\Qb{2})\cm_6(\Q{1},\gli{i},\Qb{2};;\q{3},\gli{j},\qb{4})^{\dagger})\nonumber\\
&&\hspace{0.2in} -2\re(\cm_6(\Q{1},\gli{i},\qb{4};;\q{3},\gli{j},\Qb{2})\cm_6(\Q{1},\gli{j},\Qb{2};;\q{3},\gli{i},\qb{4})^{\dagger})\nonumber\\
&&\hspace{0.2in} -2\re(\cm_6(\Q{1},\qb{4};;\q{3},\gli{i},\gli{j},\Qb{2}) \cm_6(\Q{1},\gli{j},\Qb{2};;\q{3},\gli{i},\qb{4})^{\dagger})\nonumber\\
&&\hspace{0.2in} -2\re(\cm_6(\Q{1},\gli{i},\gli{j},\qb{4};;\q{3},\Qb{2}) \cm_6(\Q{1},,\Qb{2};;\q{3},\gli{i},\gli{j},\qb{4})^{\dagger})\nonumber\\
&&\hspace{0.2in} -2\re(\cm_6(\Q{1},\qb{4};;\q{3},\gli{i},\gli{j},\Qb{2})\cm_6(\Q{1},,\Qb{2};;\q{3},\gli{i},\gli{j},\qb{4})^{\dagger})\nonumber\\
&&\hspace{0.2in} -2\re(\cm_6(\Q{1},\gli{i},\qb{4};;\q{3},\gli{j},\Qb{2})\cm_6(\Q{1},,\Qb{2};;\q{3},\gli{j},\gli{i},\qb{4})^{\dagger})\nonumber\\
&&\hspace{0.2in}+\frac{1}{N_c^2}\bigg(\frac{1}{2}|\cm_6(\Q{1},\Qb{2},\q{3},\qb{4},\phin{i},\phin{j})|^2-| \cm_6(\Q{1},\gli{i},\gli{j},\Qb{2};;\q{3},\qb{4})|^2\nonumber\\
&&\hspace{0.5in}-|\cm_6(\Q{1},\gli{i},\Qb{2};;\q{3},\gli{j},\qb{4})|^2-|\cm_6(\Q{1},\Qb{2};;\q{3},\gli{i},\gli{j},\qb{4})|^2\nonumber\\
&&\hspace{0.5in} +\re(\cm_6(\Q{1},\gli{i},\Qb{2};;\q{3},\gli{j},\qb{4})\cm_6(\Q{1},\gli{j},\Qb{2};;\q{3},\gli{i},\qb{4})^{\dagger})\nonumber\\
&&\hspace{0.5in} -\re(\cm_6(\Q{1},\gli{i},\gli{j},\Qb{2};;\q{3},\qb{4})\cm_6(\Q{1},\gli{j},\gli{i},\Qb{2};;\q{3},\qb{4})^{\dagger})\nonumber\\
&&\hspace{0.5in} -\re(\cm_6(\Q{1},\Qb{2};;\q{3},\gli{i},\gli{j},\qb{4})\cm_6(\Q{1},\Qb{2};;\q{3},\gli{j},\gli{i},\qb{4})^{\dagger})\nonumber\\
&&\hspace{0.5in} +2\re(\cm_6(\Q{1},\gli{i},\gli{j},\Qb{2};;\q{3},\qb{4}) \cm_6(\Q{1},\Qb{2};;\q{3},\gli{i},\gli{j},\qb{4})^{\dagger})\nonumber\\
&&\hspace{0.5in} +2\re(\cm_6(\Q{1},\gli{i},\gli{j},\Qb{2};;\q{3},\qb{4}) \cm_6(\Q{1},\Qb{2};;\q{3},\gli{j},\gli{i},\qb{4})^{\dagger})\bigg)\bigg]J_2^{(4)}(p_1,p_2,p_3,p_4).\nonumber\\ \label{eq.partonic2}
\eeqa
As it can be seen in eq.(\ref{eq.partonic2}), this cross section does not only contain colour-ordered matrix elements squared, but also interference terms of two different colour-ordered matrix elements.

\subsection{Subtraction terms}
In this subsection we present the subtraction terms which capture all unresolved behaviour present in the real contributions given above in eq.(\ref{eq.partonic2}) and related to the partonic process $g g \rightarrow Q \bar{Q} q \bar{q}$. As explained in section \ref{sec.formalism}, the full subtraction term receives in this case three different contributions
\beq
\ds^{S}_{g g\rightarrow Q\bar{Q}q\bar{q}}=\ds^{S,a}_{g g\rightarrow Q\bar{Q}q\bar{q}}+\ds^{S,b}_{g g\rightarrow Q\bar{Q}q\bar{q}}+\ds^{S,d}_{g g\rightarrow Q\bar{Q}q\bar{q}}.
\eeq
The $\ds^{S,a}$ terms subtract the single unresolved limits of the six-parton real radiation matrix elements. They are constructed as products of three-parton tree-level antenna functions and five-parton reduced matrix elements with remapped momenta. The $\ds^{S,b}$ pieces are genuine NNLO subtraction terms which account for those double unresolved limits of the real radiation matrix element that involve a pair of colour-connected partons; in this case these can be double soft and triple collinear limits. Finally, the $\ds^{S,d}$ terms, constructed as a product of two three-parton tree-level antenna functions and a four-parton reduced matrix element, capture the double collinear behaviour of the real matrix element squared and compensate for the over-subtraction of colour-unconnected double unresolved limits introduced in $\ds^{S,a}$. In the following, we shall outline in some detail the derivation of $\ds^{S,b}$ for which we follow a different approach than the one usually employed in the construction of antenna subtraction terms. Our starting point is the factorisation of colour sub-amplitudes in their soft $q\bar{q}$ limits.

\subsubsection{Double soft behaviour of amplitudes}
As it was explained in \cite{Abelof:2011ap} when a soft quark-antiquark pair is emitted between partons $a$ and $b$ in the colour chain, the colour-ordered amplitude denoted by 
$\cm_{n}(...,a,\qb{c};;\q{d},b,...)$
factorises as
\beq\label{eq.factqqbar}
\cm_{n}(...,a,\qb{c};;\q{d},b,...) \stackrel{^{p_c,p_d\rightarrow 0}}{\longrightarrow} [\bar{u}_{s_d}(p_d)\hspace{0.3mm}\gamma_{\mu} \hspace{0.3mm}v_{s_c}(p_c)] \left( J_a^{\mu}(p_c,p_d)-J_b^{\mu}(p_c,p_d) \right) \cm_{n-2}(...,a,b,...),
\eeq
where the soft currents are given by
\beq\label{eq.dsoftcurrent}
J_i^{\mu}(p_j,p_k)=\frac{p_i^{\mu}}{s_{jk}(s_{ij}+s_{ik})}.
\eeq
This factorisation can be applied to the first three lines of eq.(\ref{eq.colourdec}). Squaring eq.(\ref{eq.factqqbar}) and summing over the spins of the soft quark and antiquark gives
\beq
|\cm_{n}(...,a,\qb{c};;\q{d},b,...)|^2 \stackrel{^{p_c,p_d\rightarrow 0}}{\longrightarrow}\soft{a}{c}{d}{b}(m_a,m_b)|\cm_{n-2}(...,a,b,...)|^2
\eeq
with the double soft factor $\soft{a}{c}{d}{b}(m_a,m_b)$ given by
\beq\label{eq.softfactorderivation}
\soft{a}{c}{d}{b}(m_a,m_b)={\rm tr}(\not{p}_d\gamma_{\mu}\hspace{-1mm}\not{p}_c\gamma_{\nu})\left( J_a^{\mu}(p_c,p_d)-J_b^{\mu}(p_c,p_d) \right)\left( J_a^{\nu}(p_c,p_d)-J_b^{\nu}(p_c,p_d) \right)^{\dagger}.
\eeq
By explicitly evaluating the trace in eq.(\ref{eq.softfactorderivation}) and using the definition of the currents in eq.(\ref{eq.dsoftcurrent}) the expression given in eq.(\ref{eq.seik}) for the double soft factor is obtained. 

When a $q\bar{q}$ pair becomes soft in a sub-amplitude whose colour factor contains $\del{d}{c}$, like $\cm_6(\Q{1},\gli{i},\gli{j},\Qb{2};;\q{3},\qb{4})$ in the fourth line of eq.(\ref{eq.colourdec}), the factorisation given in eq.(\ref{eq.factqqbar}) does not hold. This is due to the fact that the presence of the $\del{d}{c}$ factor indicates that the gluon propagator from which the soft pair splits is photon-like. In these cases, factorisation at the amplitude level reads
\beq\label{eq.factabelian}
\cm_n(...,a;;\q{d},\qb{c})\stackrel{^{p_c,p_d\rightarrow 0}}{\longrightarrow}[\bar{u}_{s_d}(p_d)\hspace{0.3mm}\gamma_{\mu} \hspace{0.3mm}v_{p_c}(p_c)] \Bigg( \sum_{i\in\{q\}}J_i^{\mu}(p_c,p_d)-\sum_{j\in\{\bar{q}\}}J_j^{\mu}(p_c,p_d) \Bigg) \cm_{n-2}(...,a)
\eeq
where $\{q\}$ is the set of all final state quarks and initial state antiquarks in the partonic process, and $\{\bar{q}\}$ is the set of all final state antiquarks and initial state quarks. In the process $gg\to Q\bar{Q}q\bar{q}$, however, since there is only one quark-antiquark pair in addition to the soft one, each sum in eq.(\ref{eq.factabelian}) contains only one term. Also, the sub-amplitudes in the last two lines of eq.(\ref{eq.colourdec}) are finite in the soft $q\bar{q}$ limit as the quark and the antiquark are not colour connected. 

With these considerations, we can obtain the double soft behaviour of the full amplitude for the process $gg\rightarrow Q\bar{Q}q\bar{q}$ by readily applying eqs.(\ref{eq.factqqbar},\ref{eq.factabelian}) to eq.(\ref{eq.colourdec}). It reads,
\beqa
\lefteqn{M^0_6(\Q{1},\Qb{2},\q{3},\qb{4},\gli{5},\gli{6})\stackrel{^{p_3,p_4 \rightarrow0}}{\longrightarrow}g^4(\sqrt{2})^2 [\bar{u}_{s_3}(p_3)\hspace{0.3mm}\gamma_{\mu} \hspace{0.3mm}v_{s_4}(p_4)]}\nonumber\\
&&\times \sum_{(i,j)\in P(5,6)}\bigg[(T^{a_i}T^{a_j})_{i_1i_4}\delta_{i_3,i_2} \Big( J_j^{\mu}(p_4,p_3)-J_2^{\mu}(p_4,p_3)\Big)\nonumber\\
&&\hspace{0.75in}+(T^{a_i})_{i_1i_4}(T^{a_j})_{i_3 i_2} \Big( J_i^{\mu}(p_4,p_3)-J_j^{\mu}(p_4,p_3)\Big)\nonumber\\
&&\hspace{0.75in}+\delta_{i_1,i_4}(T^{a_i}T^{a_j})_{i_3 i_2} \Big( J_1^{\mu}(p_4,p_3)-J_i^{\mu}(p_4,p_3)\Big)\nonumber\\
&&\hspace{0.75in}-\frac{1}{N_c}(T^{a_i}T^{a_j})_{i_1i_2}\delta_{i_3,i_4} \Big( J_1^{\mu}(p_4,p_3)-J_2^{\mu}(p_4,p_3)\Big)\bigg]\cm_4(\Q{1},\gli{i},\gli{j},\Qb{2}).\nonumber\\ \label{eq.soft1}
\eeqa

We can now square eq.(\ref{eq.soft1}) and evaluate the Dirac traces as well as the colour traces to obtain the double soft limit of the full matrix element squared in terms of reduced colour-ordered matrix elements squared (with two partons less than the original amplitude) and combinations of double soft factors: 
\beqa
\lefteqn{|M^0_6(\Q{1},\Qb{2},\q{3},\qb{4},\gli{5},\gli{6})|^2\stackrel{^{p_3,p_4 \rightarrow0}}{\longrightarrow}g^8(N_c^2-1)}\nonumber\\
&&\times\sum_{(i,j)\in P(5,6)}\bigg[ N_c^2\Big(\soft{1}{4}{3}{i}(m_Q,0)+\soft{j}{4}{3}{2}(0,m_Q)+\soft{i}{4}{3}{j}(0,0)\Big)|\cm_4(\Q{1},\gli{i},\gli{j},\Qb{2})|^2\nonumber\\
&&\hspace{0.8in}-\soft{1}{4}{3}{2}(m_Q,m_Q)\Big(|\cm_4(\Q{1},\gli{i},\gli{j},\Qb{2})|^2-(1/2)|\cm_4(\Q{1},\phin{i},\phin{j},\Qb{2})|^2\Big)\nonumber\\
&&\hspace{0.8in}-\Big(\soft{1}{4}{3}{i}(m_Q,0)+\soft{j}{4}{3}{2}(0,m_Q)\Big)|\cm_4(\Q{1},\phin{i},\phin{j},\Qb{2})|^2\nonumber\\
&&\hspace{0.8in}+\frac{1}{2N_c^2}\soft{1}{4}{3}{2}(m_Q,m_Q)|\cm_4(\Q{1},\phin{i},\phin{j},\Qb{2})|^2\bigg].\label{eq.soft2}
\eeqa
In eq.(\ref{eq.soft2}), we see  that in the soft factors the hard radiators are two particles taken from the list $\{\Q{1},\Qb{2},\gli{i},\gli{j}\}$, and depending on which of them are involved (massive or massless) the soft factors will contain two, one, or no mass terms.

\subsection{The construction of  $\ds^{S,b}_{gg\to Q\bar{Q}q\bar{q}}$}
We decompose the $\ds^{S,b}$ subtraction term into a part which can be directly related to the double soft limit of the amplitude squared which we denote $\ds^{S,b(1)}$ and another that accounts for all other double unresolved behaviour of the real matrix element for the process $gg \to Q\bar{Q}q\bar{q}$ which is not accounted for in $\ds^{S,b(1)}$. We shall denote this latter part of the subtraction term as $\ds^{S,b(2)}$ such that, 
\begin{equation}
\ds^{S,b}_{gg\to Q\bar{Q}q\bar{q}}=\ds^{S,b(1)}_{gg\to Q\bar{Q}q\bar{q}} +\ds^{S,b(2)}_{gg\to Q\bar{Q}q\bar{q}}.
\end{equation}

From eqs.(\ref{eq.softfactb04},\ref{eq.softfactg04},\ref{eq.softfacte04}) we know that $\soft{1}{4}{3}{2}(m_Q,m_Q)$ is the double soft factor present in the double soft limit of $B_4^0(\Q{1},\qb{4},\q{3},\Qb{2})$, $\soft{1}{4}{3}{i}(m_Q,0)$ appears in the double soft limit of $E_4^0(\Q{1},\q{3},\qb{4},\gli{i})$, $\soft{2}{4}{3}{i}(m_Q,0)$ appears in the double soft limit of $E_4^0(\Qb{2},\qb{4},\q{3},\gli{i})$, and finally  $\soft{i}{4}{3}{j}(0,0)$ is the double soft limit of $G_4^0(\gli{i},\q{3},\qb{4},\gli{j})$. 

The subtraction term $\ds^{S,b(1)}$ is therefore obtained by replacing in eq.(\ref{eq.soft2}) the double soft factors with the corresponding four-parton antennae, remapping the momenta in the reduced matrix elements accordingly, and removing all spurious single collinear limits of the four-parton antennae with terms of the form $X_3^0 \cdot X_3^0$. It reads,
\beqa
\lefteqn{\ds^{S,b(1)}_{gg\rightarrow Q\bar{Q}q\bar{q}}=\norm_{LO} N_F \left( \frac{\alpha_s}{2\pi} \right)^2\frac{\bar{C}(\epsilon)^2}{C(\epsilon)^2} \dphi_4(p_1,p_2,p_3,p_4;p_5,p_6)\sum_{(i,j)\in P(5,6)} \bigg\{ } \nonumber\\
&& N_c^2 \bigg[ \Big( E_4^0(\Q{1},\q{3},\qb{4},\gli{i})-G_3^0(\gli{i},\q{3},\qb{4})D_3^0(\Q{1},\gl{(\wt{34})},\gli{\bar{i}})\nonumber \\
&&\hspace{0.5in} -A_3^0(\Q{1},\gli{i},\qb{4})E_3^0(\Q{(\wt{14})},\q{3},\qi{\bar{i}})\Big)|\cm_4(\Q{(\wt{134})},\gli{\bar{i}},\gli{j},\Qb{2})|^2 J_2^{(2)}(\wt{p_{134}},p_2)\nonumber\\
&&\hspace{0.17in} +\Big( E_4^0(\Qb{2},\qb{4},\q{3},\gli{i})-G_3^0(\gli{i},\q{3},\qb{4})D_3^0(\Qb{2},\gl{(\wt{34})},\gli{\bar{i}})\nonumber \\
&&\hspace{0.5in} -A_3^0(\Qb{2},\gli{i},\qb{3})E_3^0(\Qb{(\wt{23})},\qb{4},\qbi{\bar{i}})\Big)|\cm_4(\Q{1},\gli{j},\gli{\bar{i}},\Qb{(\wt{234})})|^2 J_2^{(2)}(p_1,\wt{p_{234}})\nonumber\\
&&\hspace{0.17in} +\Big(G_4^0(\gli{i},\q{3},\qb{4},\gli{j})-(1/2)G_3^0(\gli{i},\q{3},\qb{4})F_3^0(\gli{\bar{i}},\gl{(\wt{34})},\gli{j})\nonumber \\
&&\hspace{0.5in} -(1/2)G_3^0(\gli{j},\q{3},\qb{4})F_3^0(\gli{i},\gl{(\wt{34})},\gli{\bar{j}})-a_3^0(\q{3},\gli{i},\qb{4})G_3^0(\gli{j},\qb{(\wt{34})},\qbi{\bar{i}})\nonumber\\
&&\hspace{0.5in} -a_3^0(\qb{4},\gli{j},\q{3})G_3^0(\gli{i},\q{(\wt{34})},\qi{\bar{j}})\Big) |\cm_4(\Q{\tilde{1}},\gli{\bar{i}},\gli{\bar{j}},\Qb{\tilde{2}})|^2 J_2^{(2)}(\tilde{p_1},\tilde{p_2})\bigg]\nonumber\\
&& -\Big( E_4^0(\Q{1},\q{3},\qb{4},\gli{i})-G_3^0(\gli{i},\q{3},\qb{4})D_3^0(\Q{1},\gl{(\wt{34})},\gli{\bar{i}})\nonumber \\
&&\hspace{0.35in} -A_3^0(\Q{1},\gli{i},\qb{4})E_3^0(\Q{(\wt{14})},\q{3},\qi{\bar{i}})\Big)|\cm_4(\Q{(\wt{134})},\phin{\bar{i}},\phin{j},\Qb{2})|^2 J_2^{(2)}(\wt{p_{134}},p_2)\nonumber\\
&& -\Big( E_4^0(\Qb{2},\qb{4},\q{3},\gli{i})-G_3^0(\gli{i},\q{3},\qb{4})D_3^0(\Qb{2},\gl{(\wt{34})},\gli{\bar{i}})\nonumber \\
&&\hspace{0.35in} -A_3^0(\Qb{2},\gli{i},\qb{3})E_3^0(\Qb{(\wt{23})},\qb{4},\qbi{\bar{i}})\Big)|\cm_4(\Q{1},\phin{j},\phin{\bar{i}},\Qb{(\wt{234})})|^2 J_2^{(2)}(p_1,\wt{p_{234}})\nonumber\\
&& -\Big( B_4^0(\Q{1},\qb{4},\q{3},\Qb{2})  -(1/2)E_3^0(\Q{1},\q{3},\qb{4})A_3^0(\Q{(\wt{14})},\gl{(\wt{34})},\Qb{2}) \nonumber\\
&&\hspace{0.35in} -(1/2)E_3^0(\Q{2},\q{3},\qb{4})A_3^0(\Q{1},\gl{(\wt{34})},\Qb{(\wt{24})}) \Big)\times \Big( |\cm_4(\Q{(\wt{134})},\gli{i},\gli{j},\Qb{(\wt{234})}|^2\nonumber\\
&&\hspace{0.35in}  -(1/2) |\cm_4(\Q{(\wt{134})},\phin{i},\phin{j},\Qb{(\wt{234})}|^2\Big) J_2^{(2)}(\wt{p_{134}},\wt{p_{234}})\nonumber\\
&&+\frac{1}{N_c^2}\bigg[\Big( B_4^0(\Q{1},\qb{4},\q{3},\Qb{2}) -(1/2)E_3^0(\Q{1},\q{3},\qb{4})A_3^0(\Q{(\wt{14})},\gl{(\wt{34})},\Qb{2}) \nonumber\\
&&\hspace{0.65in} -(1/2)E_3^0(\Q{2},\q{3},\qb{4})A_3^0(\Q{1},\gl{(\wt{34})},\Qb{(\wt{24})}) \Big)\nonumber\\
&&\hspace{0.75in} \times |\cm_4(\Q{(\wt{134})},\phin{i},\phin{j},\Qb{(\wt{234})}|^2 J_2^{(2)}(\wt{p_{134}},\wt{p_{234}})\bigg]\bigg\}. \label{eq.subtbsoft}
\eeqa

As explained in section \ref{sec.limits}, the initial-final massive $D_3^0$ antenna appearing in this subtraction term, is  the ``full'' antenna. It is this antenna onto which the $E_4^0$ antenna function collapses to in its single $q\bar{q}$ collinear limit.

In order to obtain the full $\ds^{S,b}$ subtraction term, we have to supplement eq.(\ref{eq.subtbsoft}) with additional terms that ensure the correct subtraction of all other colour-connected double unresolved limit of the real radiation matrix element squared. The only other unresolved limits of this type that the process $gg\to Q\bar{Q}q\bar{q}$ has are triple collinear limits $\gli{i}||\q{3}||\q{4}$. In these limits the full matrix element squared factorises as \cite{Campbell:1997hg,Catani:1998nv,Catani:1999ss,deFlorian:2001zd}
\beqa
&&\hspace{-0.3in} |M^0_6(\Q{1},\Qb{2},\q{3},\qb{4},\gli{5},\gli{6})|^2\stackrel{^{\gli{i}||\q{3}||\q{4}}}{\longrightarrow}g^2\bigg[ N_c \left(P_{g_i \bar{q}_4 q_3\leftarrow G}+P_{\bar{q}_4 q_3 g_i\leftarrow G} \right) \nonumber\\
&&\hspace{2.3in}-\frac{1}{N_c}P_{q_3 g_i \bar{q}_4\leftarrow G} \bigg]|M^0_4(\Q{1},\Qb{2},\widehat{(34i)}_g,\gli{j})|^2.\label{eq.tc1}
\eeqa
Taking now the triple collinear limit of $\ds^{S,b(1)}$ we obtain, 
\beqa
\lefteqn{\ds^{S,b(1)}_{g\bar{g}\rightarrow Q\bar{Q}q\bar{q}} \stackrel{^{\gli{i}||\q{3}||\q{4}}}{\longrightarrow}\norm_{LO} N_F \left( \frac{\alpha_s}{2\pi} \right)^2\frac{\bar{C}(\epsilon)^2}{C(\epsilon)^2} \dphi_4(p_1,p_2,p_3,p_4;p_5,p_6)}\nonumber\\
&&\hspace{1.25in} \times N_c \left(P_{g_i \bar{q}_4 q_3\leftarrow G}+P_{\bar{q}_4 q_3 g_i\leftarrow G} \right) |M^0_4(\Q{1},\Qb{2},\widehat{(34i)}_g,\gli{j})|^2 J_2^{(2)}(p_1,p_2).\nonumber\\ \label{eq.tc2}
\eeqa
where we have used eqs.(\ref{eq.cdec4},\ref{eq.cdec42}) to relate the full four-parton reduced matrix element squared denoted by $|M^0_4(\Q{1},\Qb{2},\widehat{(34i)}_g,\gli{j})|^2$ to the corresponding colour-ordered amplitudes. The triple collinear splitting functions $P_{ijk \leftarrow G}$ have been defined in section~\ref{sec.limits}. For conciseness, in the above equations the arguments of these splitting functions have been omitted. 

We see that the subtraction term $\ds^{S,b(1)}$, originally derived in order to capture the soft $q\bar{q}$ limit, also subtracts the non-abelian piece of the triple collinear limits $\gli{i}||\q{3}||\q{4}$. To account for the QED-like triple collinear limits, proportional to $1/N_c$ in eq.(\ref{eq.tc1}), we use initial-final $\wt{E}_4^0$ massive four-parton antennae. The following subtraction term is obtained:
\beqa
\lefteqn{\ds^{S,b(2)}_{g g\rightarrow Q\bar{Q}q\bar{q}}=\norm_{LO} N_F \left( \frac{\alpha_s}{2\pi} \right)^2\frac{\bar{C}(\epsilon)^2}{C(\epsilon)^2} \dphi_4(p_1,p_2,p_3,p_4;p_5,p_6)\sum_{(i,j)\in P(5,6)} \bigg\{ } \nonumber\\
&&-\frac{1}{2}\Big( \wt{E}_4^0(\Q{1},\q{3},\qb{4},\gli{i})-A_3^0(\Q{1},\gli{i},\qb{4})E_3^0(\Q{(\wt{14})},\q{3},\qi{\bar{i}}) \nonumber\\
&&\hspace{0.35in} -A_3^0(\Q{1},\gli{i},\q{3})E_3^0(\Q{(\wt{13})},\qb{4},\qbi{\bar{i}})\Big)\times\Big(|\cm_4(\Q{(\wt{134})},\gli{\bar{i}},\gli{j},\Qb{2}|^2\nonumber\\
&&\hspace{0.35in} +|\cm_4(\Q{(\wt{134})},\gli{j},\gli{\bar{i}},\Qb{2}|^2\Big) J_2^{(2)}(\wt{p_{134}},p_2)\nonumber\\
&&-\frac{1}{2}\Big( \wt{E}_4^0(\Qb{2},\qb{4},\q{3},\gli{i})-A_3^0(\Qb{2},\gli{i},\q{3})E_3^0(\Qb{(\wt{23})},\qb{4},\qbi{\bar{i}}) \nonumber\\
&&\hspace{0.35in} -A_3^0(\Qb{2},\gli{i},\qb{4})E_3^0(\Q{(\wt{24})},\q{3},\qi{\bar{i}})\Big)\times\Big(|\cm_4(\Q{1},\gli{\bar{i}},\gli{j},\Qb{(\wt{234})}|^2\nonumber\\
&&\hspace{0.35in} +|\cm_4(\Q{1},\gli{j},\gli{\bar{i}},\Qb{(\wt{234})}|^2\Big) J_2^{(2)}(p_1,\wt{p_{234}})\nonumber\\
&&+\frac{1}{N_c^2}\bigg[\frac{1}{2}\Big( \wt{E}_4^0(\Q{1},\q{3},\qb{4},\gli{i})-A_3^0(\Q{1},\gli{i},\qb{4})E_3^0(\Q{(\wt{14})},\q{3},\qi{\bar{i}})\nonumber\\
&&\hspace{0.65in} -A_3^0(\Q{1},\gli{i},\q{3})E_3^0(\Q{(\wt{13})},\qb{4},\qbi{\bar{i}})\Big)|\cm_4(\Q{(\wt{134})},\phin{\bar{i}},\phin{j},\Qb{2}|^2 J_2^{(2)}(\wt{p_{134}},p_2)\nonumber\\
&&\hspace{0.32in} +\frac{1}{2}\Big( \wt{E}_4^0(\Qb{2},\qb{4},\q{3},\gli{i})-A_3^0(\Qb{2},\gli{i},\q{3})E_3^0(\Qb{(\wt{23})},\qb{4},\qbi{\bar{i}}) \nonumber\\
&&\hspace{0.65in} -A_3^0(\Qb{2},\gli{i},\qb{4})E_3^0(\Q{(\wt{24})},\q{3},\qi{\bar{i}})\Big)|\cm_4(\Q{1},\phin{\bar{i}},\phin{j},\Qb{(\wt{234})}|^2J_2^{(2)}(p_1,\wt{p_{234}})\bigg]\bigg\}.\nonumber\\ \label{eq.subtbtc}
\eeqa
Since the $\wt{E}_4^0$ antenna does not possess any soft $q\bar{q}$ limit, the subtraction term given in eq.(\ref{eq.subtbtc}) does not have any soft $q\bar{q}$ limits either and only subtracts the abelian triple collinear limit, as required. The sum $\ds^{S,b(1)}+\ds^{S,b(2)}$ will therefore correctly subtract both double soft and triple collinear limits of the double real radiation matrix element squared, without introducing any spurious single unresolved singularities.

\subsection{Construction of the $\ds^{S,a}_{gg\to Q\bar{Q}q\bar{q}}$ and $\ds^{S,d}_{gg\to Q\bar{Q}q\bar{q}}$ subtraction terms}
The $\ds^{S,a}$ subtraction terms are NLO like: They are constructed as products of three-parton tree-level antennae and five-parton reduced matrix elements, and they subtract the single unresolved limits of the real radiation matrix elements. Thus, for the present calculation, they could a priori be taken over from \cite{Abelof:2011jv}, where we derived the NLO subtraction terms for $t\bar{t}+jet$ production. 

However, at NNLO, the five-parton reduced matrix elements with remapped momenta present in these NLO-like subtraction terms develop further single unresolved limits. Those singularities are unphysical, since they do not correspond to any unresolved behaviour of the real radiation matrix element squared and must be cancelled by the $X_3^0\cdot X_3^0$ pieces of the b-type subtraction terms. In $\ds^{S,b}$, these $X_3^0\cdot X_3^0$ terms are dictated by the requirement that all single unresolved limits of the four-parton antennae should be removed. As the choice of these four-parton antennae is fixed by the double unresolved behaviour of the real radiation matrix element, we do not have the freedom to choose the three-parton antenna functions that make $\ds^{S,a}$ simplest, as we did in \cite{Abelof:2011jv}. Instead, these three-parton antennae have to be carefully chosen in such a way that the cancelation of the aforementioned unphysical singularities is achieved. With these considerations we obtain the $\ds^{S,a}$ subtraction term as
\beqa
\lefteqn{\ds^{S,a}_{g g\rightarrow Q\bar{Q}q\bar{q}}=\norm_{LO} N_F \left( \frac{\alpha_s}{2\pi} \right)^2\frac{\bar{C}(\epsilon)^2}{C(\epsilon)^2} \dphi_4(p_1,p_2,p_3,p_4;p_5,p_6)\sum_{(i,j)\in P(5,6)} \bigg\{  } \nonumber\\
&&N_c^2 \bigg[ G_3^0(\gli{i},\q{3},\qb{4})\Big( |\cm_5(\Q{1},\gl{(\wt{34})},\gli{\bar{i}},\gli{j},\Qb{2})|^2 +|\cm_5(\Q{1},\gli{j},\gli{\bar{i}},\gl{(\wt{34})},\Qb{2})|^2\nonumber\\
&&\hspace{0.5in} +(1/2)\big(|\cm_5(\Q{1},\gli{\bar{i}},\gl{(\wt{34})},\gli{j},\Qb{2})|^2 +|\cm_5(\Q{1},\gli{j},\gl{(\wt{34})},\gli{\bar{i}},\Qb{2})|^2\big)\Big) J_2^{(3)}(p_1,p_2,\wt{p_{34}})\nonumber\\
&&\hspace{0.17in} +A_3^0(\Qb{2},\gli{i},\q{3})|\cm_5(\Q{1},\gli{j},\qb{4};;\qbi{\bar{i}},\Qb{(\wt{23})})|^2 J_2^{(3)}(p_1,\wt{p_{23}},p_4)\nonumber\\
&&\hspace{0.17in} +a_3^0(\q{3},\gli{i},\qb{4})|\cm_5(\Q{1},\qb{(\wt{34})};;\qbi{\bar{i}},\gli{j},\Qb{2})|^2 J_2^{(3)}(p_1,p_2,\wt{p_{34}})\nonumber\\
&&\hspace{0.17in} +A_3^0(\Q{1},\gli{i},\qb{4})|\cm_5(\Q{(\wt{14})},\qi{\bar{i}};;\q{3},\gli{j},\Qb{2})|^2 J_2^{(3)}(\wt{p_{14}},p_2,p_3)\nonumber\\
&&\hspace{0.17in} +a_3^0(\qb{4},\gli{i},\q{3})|\cm_5(\Q{1},\gli{j},\qi{\bar{i}};;\q{(\wt{34})},\Qb{2})|^2 J_2^{(3)}(p_1,p_2,\wt{p_{34}})\bigg]\nonumber\\
&&-G_3^0(\gli{i},\q{3},\qb{4})\Big( |\cm_5(\Q{1},\gl{(\wt{34})},\gli{\bar{i}},\phin{j},\Qb{2})|^2+|\cm_5(\Q{1},\gli{\bar{i}},\gl{(\wt{34})},\phin{j},\Qb{2})|^2\Big) J_2^{(3)}(p_1,p_2,\wt{p_{34}}) \nonumber\\
&&-\frac{1}{2}E_3^0(\Q{1},\q{3},\qb{4})|\cm_5(\Q{(\wt{13})},\gli{i},\gli{j},\ph{(\wt{34})},\Qb{2})|^2 J_2^{(3)}(\wt{p_{13}},p_2,\wt{p_{34}})\nonumber\\
&&-\frac{1}{2}E_3^0(\Qb{2},\q{3},\qb{4})|\cm_5(\Q{1},\gli{i},\gli{j},\ph{(\wt{34})},\Qb{(\wt{23})})|^2 J_2^{(3)}(p_1,\wt{p_{23}},\wt{p_{34}})\nonumber\\
&&-\frac{1}{2}A_3^0(\Q{1},\gli{i},\q{3})\Big( |\cm_5(\Q{(\wt{13})},\gli{j},\qb{4};;\qbi{\bar{i}},\Qb{2})|^2\nonumber\\
&&\hspace{0.35in} +|\cm_5(\Q{(\wt{13})},\qb{4};;\qbi{\bar{i}},\gli{j},\Qb{2})|^2\Big) J_2^{(3)}(\wt{p_{13}},p_2,p_4) \nonumber\\
&&-\frac{1}{2}A_3^0(\Qb{2},\gli{i},\qb{4})\Big( |\cm_5(\Q{1},\qi{\bar{i}};;\q{3},\gli{j},\Qb{(\wt{24})})|^2\nonumber\\
&&\hspace{0.35in} +|\cm_5(\Q{1},\gli{j},\qi{\bar{i}};;\q{3},\Qb{(\wt{24})})|^2\Big) J_2^{(3)}(p_1,\wt{p_{23}},p_3) \nonumber\\
&&+A_3^0(\Q{1},\gli{i},\qb{4})\Big( |\cm_5(\Q{(\wt{14})},\Qb{2};;\q{3},\gli{j},\qi{\bar{i}})|^2+|\cm_5(\Q{(\wt{14})},\gli{j},\Qb{2};;\q{3},\qi{\bar{i}})|^2 \nonumber\\
&&\hspace{0.35in}-2|\cm_5(\Q{(\wt{14})},\Qb{2},\q{3},\qi{\bar{i}},\phin{j})|^2-(1/2)|\cm_5(\Q{(\wt{14})},\qi{\bar{i}};;\q{3},\gli{j},\Qb{2})|^2 \nonumber\\
&&\hspace{0.35in}-(1/2)|\cm_5(\Q{(\wt{14})},\gli{j},\qi{\bar{i}};;\q{3},\Qb{2})|^2\Big) J_2^{(3)}(\wt{p_{14}},p_2,p_3) \nonumber\\
&&+A_3^0(\Q{2},\gli{i},\qb{3})\Big( |\cm_5(\Q{1},\Qb{(\wt{23})};;\qbi{\bar{i}},\gli{j},\qb{4})|^2+|\cm_5(\Q{1},\gli{j},\Qb{(\wt{23})};;\qbi{\bar{i}},\qb{4})|^2 \nonumber\\
&&\hspace{0.35in} -2|\cm_5(\Q{1},\Qb{(\wt{23})},\qbi{\bar{i}},\qb{4},\phin{j})|^2-(1/2)|\cm_5(\Q{(1},\qb{4};;\qbi{\bar{i}},\gli{j},\Qb{(\wt{23})})|^2\nonumber\\
&&\hspace{0.35in}-(1/2)|\cm_5(\Q{(1},\gli{j},\qb{4};;\qbi{\bar{i}},\Qb{(\wt{23})})|^2\Big) J_2^{(3)}(p_1,\wt{p_{23}},p_4) \nonumber\\
&&+\frac{1}{N_c^2}\bigg[\frac{1}{2}E_3^0(\Q{1},\q{3},\qb{4})|\cm_5(\Q{(\wt{13})},\phin{i},\phin{j},\ph{(\wt{34})},\Qb{2})|^2 J_2^{(3)}(\wt{p_{13}},p_2,\wt{p_{34}})\nonumber\\
&&\hspace{0.32in}+\frac{1}{2}E_3^0(\Qb{2},\q{3},\qb{4})|\cm_5(\Q{1},\phin{i},\phin{j},\ph{(\wt{34})},\Qb{(\wt{23})})|^2 J_2^{(3)}(p_1,\wt{p_{23}},\wt{p_{34}})\nonumber\\
&&\hspace{0.32in}-\frac{1}{2}A_3^0(\Q{1},\gli{i},\qb{4})\Big( |\cm_5(\Q{(\wt{14})},\Qb{2};;\q{3},\gli{j},\qi{\bar{i}})|^2+|\cm_5(\Q{(\wt{14})},\gli{j},\Qb{2};;\q{3},\qi{\bar{i}})|^2 \nonumber\\
&&\hspace{0.65in}-2|\cm_5(\Q{(\wt{14})},\Qb{2},\q{3},\qi{\bar{i}},\phin{j})|^2\Big)J_2^{(3)}(\wt{p_{14}},p_2,p_3) \nonumber\\
&&\hspace{0.32in} -\frac{1}{2}A_3^0(\Q{1},\gli{i},\q{3})\Big( |\cm_5(\Q{(\wt{13})},\Qb{2};;\qbi{\bar{i}},\gli{j},\qb{4})|^2+|\cm_5(\Q{(\wt{13})},\gli{j},\Qb{2};;\qbi{\bar{i}},\qb{4})|^2 \nonumber\\
&&\hspace{0.65in}-2|\cm_5(\Q{(\wt{13})},\Qb{2},\qbi{\bar{i}},\qb{4},\phin{j})|^2\Big)J_2^{(3)}(\wt{p_{13}},p_2,p_4) \nonumber\\
&&\hspace{0.32in} -\frac{1}{2}A_3^0(\Qb{2},\gli{i},\q{3})\Big( |\cm_5(\Q{1},\Qb{(\wt{23})};;\qbi{\bar{i}},\gli{j},\qb{4})|^2+|\cm_5(\Q{1},\gli{j},\Qb{(\wt{23})};;\qbi{\bar{i}},\qb{4})|^2 \nonumber\\
&&\hspace{0.65in}-2|\cm_5(\Q{1},\Qb{(\wt{23})},\qbi{\bar{i}},\qb{4},\phin{j})|^2\Big)J_2^{(3)}(\wt{p_{13}},p_2,p_4) \nonumber\\
&&\hspace{0.32in} -\frac{1}{2}A_3^0(\Qb{2},\gli{i},\qb{4})\Big( |\cm_5(\Q{1},\Qb{(\wt{24})};;\q{3},\gli{j},\qi{\bar{i}})|^2+|\cm_5(\Q{1},\gli{j},\Qb{(\wt{24})};;\q{3},\qi{\bar{i}})|^2 \nonumber\\
&&\hspace{0.65in}-2|\cm_5(\Q{1},\Qb{(\wt{24})},\q{3},\qi{\bar{i}},\phin{j})|^2\Big)J_2^{(3)}(p_1,\wt{p_{24}},p_3) \bigg]\bigg\}.\nonumber\\ \label{eq.subta}
\eeqa

Finally, the $\ds^{S,d}$ terms are constructed in the usual fashion. They capture the double unresolved behaviour of the real matrix element squared when two colour-unconnected unresolved partons are present and they remove the double counting of double unresolved limits in $\ds^{S,a}$. They are given by
\beqa
\lefteqn{\ds^{S,d}_{g\bar{g}\rightarrow Q\bar{Q}q\bar{q}}=-\norm_{LO} N_F \left( \frac{\alpha_s}{2\pi} \right)^2\frac{\bar{C}(\epsilon)^2}{C(\epsilon)^2} \dphi_4(p_1,p_2,p_3,p_4;p_5,p_6)\times} \nonumber\\
&&\sum_{(i,j)\in P(5,6)} \bigg\{ N_c^2 A_3^0(\Q{1},\gli{i},\qb{4})A_3^0(\Qb{2},\gli{j},\q{3})|\cm_4(\Q{(\wt{14})},\Qb{(\wt{23})},\qbi{\bar{j}},\qi{\bar{i}})|^2 J_2^{(2)}(\wt{p_{14}},\wt{p_{23}})\nonumber\\
&&\hspace{21mm} -\frac{3}{2}A_3^0(\Q{1},\gli{i},\qb{4})A_3^0(\Qb{2},\gli{j},\q{3})|\cm_4(\Q{(\wt{14})},\Qb{(\wt{23})},\qbi{\bar{j}},\qi{\bar{i}})|^2 J_2^{(2)}(\wt{p_{14}},\wt{p_{23}})\nonumber\\
&&\hspace{21mm} -\frac{1}{2}A_3^0(\Q{1},\gli{i},\q{3})A_3^0(\Qb{2},\gli{i},\qb{4})|\cm_4(\Q{(\wt{13})},\Qb{(\wt{24})},\qbi{\bar{i}},\qi{\bar{j}})|^2 J_2^{(2)}(\wt{p_{13}},\wt{p_{24}})\nonumber\\
&&\hspace{18mm} +\frac{1}{N_c^2}\bigg[\frac{1}{2}A_3^0(\Q{1},\gli{i},\qb{4})A_3^0(\Qb{2},\gli{j},\q{3})|\cm_4(\Q{(\wt{14})},\Qb{(\wt{23})},\qbi{\bar{j}},\qi{\bar{i}})|^2 J_2^{(2)}(\wt{p_{14}},\wt{p_{23}})\nonumber\\
&&\hspace{26mm} +\frac{1}{2}A_3^0(\Q{1},\gli{i},\q{3})A_3^0(\Qb{2},\gli{i},\qb{4})|\cm_4(\Q{(\wt{13})},\Qb{(\wt{24})},\qbi{\bar{i}},\qi{\bar{j}})|^2 J_2^{(2)}(\wt{p_{13}},\wt{p_{24}})\bigg]\bigg\}.\nonumber\\ \label{eq.subtd}
\eeqa

\section{Numerical results}\label{sec.results}
To verify how well the subtraction terms approximate the double real contributions related to the partonic process $gg \rightarrow Q\bar{Q} q \bar{q}$, we have used {\tt RAMBO}~\cite{Kleiss:1985gy} to generate phase space points in the vicinity of the singular regions and computed the ratio
\beq
R=\frac{\ds^{RR}_{NNLO}}{\ds^S_{NNLO}}
\eeq
for each of these points. As before, $\ds^{RR}_{NNLO}$ stands for the double real radiation contributions while $\ds^S_{NNLO}$ is the corresponding subtraction term. In each unresolved limit we define a control variable $x$ that allows us to vary the proximity of the phase space points to the singularity. For the difference $\ds^{RR}_{NNLO}-\ds^S_{NNLO}$ to be finite and numerically integrable in four dimensions, the ratio $R$ should approach unity as we get close to any singularity. The phase space points were generated with a fixed centre-of-mass energy of $\sqrt{s}=1000$ GeV, the heavy fermions were given a mass of $174.3$ GeV, and the two hard jets were required to have $p_T > 50$ GeV.

\subsection{Double soft limit}
\begin{figure}[ht]
\centering
\subfigure[]{
\scalebox{0.8}{
\setlength{\unitlength}{4144sp}

\begingroup\makeatletter\ifx\SetFigFont\undefined
\gdef\SetFigFont#1#2#3#4#5{
  \reset@font\fontsize{#1}{#2pt}
  \fontfamily{#3}\fontseries{#4}\fontshape{#5}
  \selectfont}
\fi\endgroup
\begin{picture}(3296,2656)(1576,-2585)

\thicklines
{\put(3021,-1323){\vector(-1,-3){364.500}}
}
{\put(4321,-1276){\vector(-1, 0){1170}}
}
{\put(3026,-1235){\vector( 1, 3){364.500}}
}

\thinlines
{\put(2832,-1754){\vector(-4,-1){222.353}}
}
{\put(3196,-871){\vector( 4, 1){222.353}}

\put(3200,-61){\makebox(0,0)[lb]{\smash{{\SetFigFont{12}{14.4}{\rmdefault}{\mddefault}{\updefault}{$1_Q$}
}}}}
\put(2350,-2550){\makebox(0,0)[lb]{\smash{{\SetFigFont{12}{14.4}{\rmdefault}{\mddefault}{\updefault}{$2_{\bar{Q}}$}
}}}}
}
\put(1370,-1350){\makebox(0,0)[lb]{\smash{{\SetFigFont{12}{14.4}{\rmdefault}{\mddefault}{\updefault}{$5_g$}
}}}}
\put(4200,-1321){\makebox(0,0)[lb]{\smash{{\SetFigFont{12}{14.4}{\rmdefault}{\mddefault}{\updefault}{$6_g$}
}}}}
\put(3300,-880){\makebox(0,0)[lb]{\smash{{\SetFigFont{12}{14.4}{\rmdefault}{\mddefault}{\updefault}{$3_q$}
}}}}
\put(2170,-1861){\makebox(0,0)[lb]{\smash{{\SetFigFont{12}{14.4}{\rmdefault}{\mddefault}{\updefault}{$4_{\bar{q}}$}
}}}}
\thicklines
{\put(1756,-1276){\vector( 1, 0){1170}}
}
\end{picture}
\label{fig.pic1}
}}
\subfigure[]{
\includegraphics[scale=0.6]{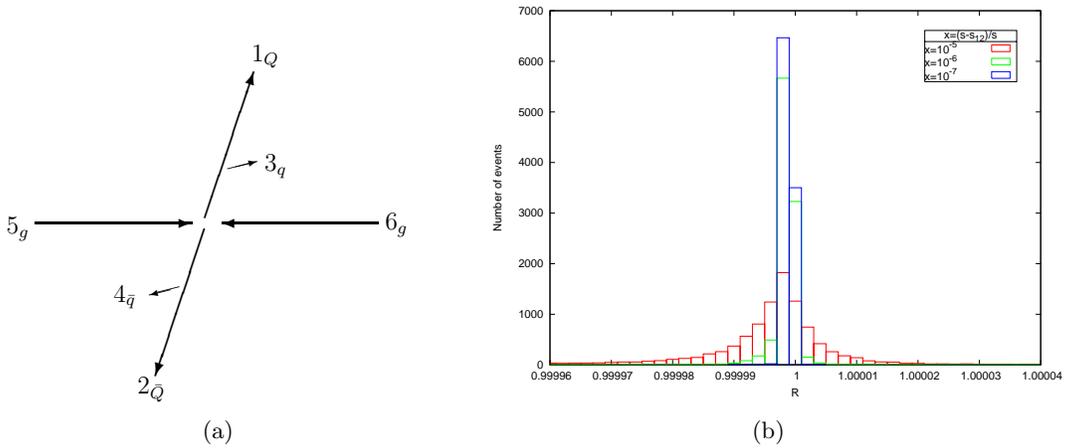}
\label{fig.ds}
}
\caption[]{\subref{fig.pic1} Ilustration of a double soft $q \bar{q}$ event. \subref{fig.ds} Distribution of R for 10000 double soft phase space points.}
\end{figure}

The double soft phase space configurations are characterised by the $Q\bar{Q}$ pair taking nearly the full center of mass energy of the event $s$, leaving the massless final state $q\bar{q}$ pair with almost zero energy, as depicted in fig.\ref{fig.pic1}.
In fig.\ref{fig.ds} we show the ratio between the double real
radiation matrix element and the subtraction term for three different
values of the control variable $x$ defined in this case by, $x=(s-s_{12})/s$. It can be seen that as the soft $q\bar{q}$ pair, takes a smaller share of the total energy, i.e. as $x$ becomes smaller, the peak of the distribution around $R=1$ is sharper. This is a sign that the approximation improves as the limit is approached.

\subsection{Triple collinear limit}
\begin{figure}[ht]
\centering
\subfigure[]{
\scalebox{0.8}{
\setlength{\unitlength}{4144sp}

\begingroup\makeatletter\ifx\SetFigFont\undefined
\gdef\SetFigFont#1#2#3#4#5{
  \reset@font\fontsize{#1}{#2pt}
  \fontfamily{#3}\fontseries{#4}\fontshape{#5}
  \selectfont}
\fi\endgroup
\begin{picture}(2927,2640)(1576,-2581)
\thicklines
{\put(3021,-1323){\vector(-1,-3){364.500}}
}
{\put(4321,-1276){\vector(-1, 0){1170}}
}
{\put(3026,-1235){\vector( 1, 3){364.500}}
}
{\put(1910,-1243){\vector( 4, 1){641.412}}
}
{\put(1911,-1304){\vector( 4,-1){684}}
}
{\put(1756,-1276){\vector( 1, 0){1170}}
}
\put(1340,-1340){\makebox(0,0)[lb]{\smash{{\SetFigFont{12}{14.4}{\rmdefault}{\mddefault}{\updefault}{$5_g$}
}}}}
\put(4230,-1321){\makebox(0,0)[lb]{\smash{{\SetFigFont{12}{14.4}{\rmdefault}{\mddefault}{\updefault}{$6_g$}
}}}}
\put(2415,-2600){\makebox(0,0)[lb]{\smash{{\SetFigFont{12}{14.4}{\rmdefault}{\mddefault}{\updefault}{$2_{\bar{Q}}$}
}}}}
\put(3200,-61){\makebox(0,0)[lb]{\smash{{\SetFigFont{12}{14.4}{\rmdefault}{\mddefault}{\updefault}{$1_Q$}
}}}}
\put(2490,-1100){\makebox(0,0)[lb]{\smash{{\SetFigFont{12}{14.4}{\rmdefault}{\mddefault}{\updefault}{$3_q$}
}}}}
\put(2490,-1550){\makebox(0,0)[lb]{\smash{{\SetFigFont{12}{14.4}{\rmdefault}{\mddefault}{\updefault}{$4_{\bar{q}}$}
}}}}
\end{picture}
\label{fig.pic2}
}}
\subfigure[]{
\includegraphics[scale=0.6]{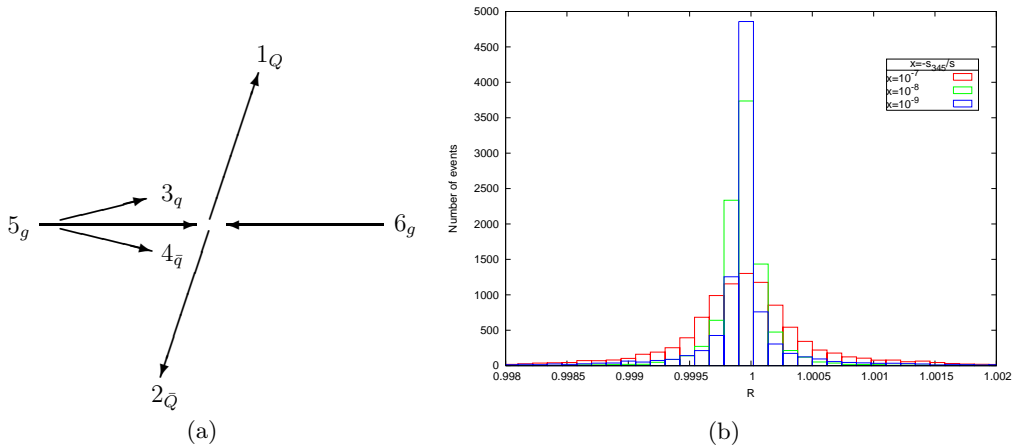}
\label{fig.tc}
}
\caption[]{\subref{fig.pic1} Ilustration of a triple collinear event. \subref{fig.ds} Distribution of R for 10000 triple collinear phase space points.}
\end{figure}

Other double unresolved configurations in which the amplitude for this process is singular are the triple collinear limits $\gli{i}||\q{3}||\qb{4}$ illustrated in fig.\ref{fig.pic2}.
In fig.\ref{fig.tc} we show how, as we make the control variable $x=-s_{345}/s$ smaller, that is, as we get closer in phase space to the singularity $\gli{5}||\q{3}||\qb{4}$ of the real radiation matrix element squared, the histogram becomes more pronouncedly peaked around unity. This signals us again that the approximation is good. Similar results which are not shown are obtained for the triple collinear limit involving the other initial state gluon $\gli{6}$.

\subsection{Double collinear limits}
The last type of double unresolved limits that the amplitude for the process $gg\rightarrow Q\bar{Q}q\bar{q}$ has, are the initial-final double collinear limits $\gli{i}||\q{3}+\gli{j}||\qb{4}$.
\begin{figure}[ht]
\centering
\subfigure[]{
\scalebox{0.8}{
\setlength{\unitlength}{4144sp}

\begingroup\makeatletter\ifx\SetFigFont\undefined
\gdef\SetFigFont#1#2#3#4#5{
  \reset@font\fontsize{#1}{#2pt}
  \fontfamily{#3}\fontseries{#4}\fontshape{#5}
  \selectfont}
\fi\endgroup
\begin{picture}(2927,2640)(1576,-2581)

\thicklines
{\put(3021,-1323){\vector(-1,-3){364.500}}
}
{\put(1756,-1276){\vector( 1, 0){1170}}
}
{\put(4321,-1276){\vector(-1, 0){1170}}
}
{\put(3026,-1235){\vector( 1, 3){364.500}}
}
{\put(1910,-1243){\vector( 4, 1){641.412}}
}
{\put(4212,-1327){\vector(-4,-1){641.412}}
}
\put(1340,-1330){\makebox(0,0)[lb]{\smash{{\SetFigFont{12}{14.4}{\rmdefault}{\mddefault}{\updefault}{$5_g$}
}}}}
\put(4270,-1321){\makebox(0,0)[lb]{\smash{{\SetFigFont{12}{14.4}{\rmdefault}{\mddefault}{\updefault}{$6_g$}
}}}}
\put(2460,-1110){\makebox(0,0)[lb]{\smash{{\SetFigFont{12}{14.4}{\rmdefault}{\mddefault}{\updefault}{$3_q$}
}}}}
\put(2400,-2630){\makebox(0,0)[lb]{\smash{{\SetFigFont{12}{14.4}{\rmdefault}{\mddefault}{\updefault}{$2_{\bar{Q}}$}
}}}}
\put(3230,-61){\makebox(0,0)[lb]{\smash{{\SetFigFont{12}{14.4}{\rmdefault}{\mddefault}{\updefault}{$1_Q$}
}}}}
\put(3180,-1530){\makebox(0,0)[lb]{\smash{{\SetFigFont{12}{14.4}{\rmdefault}{\mddefault}{\updefault}{$4_{\bar{q}}$}
}}}}

\end{picture}
\label{fig.pic3}
}}
\subfigure[]{
\includegraphics[scale=0.6]{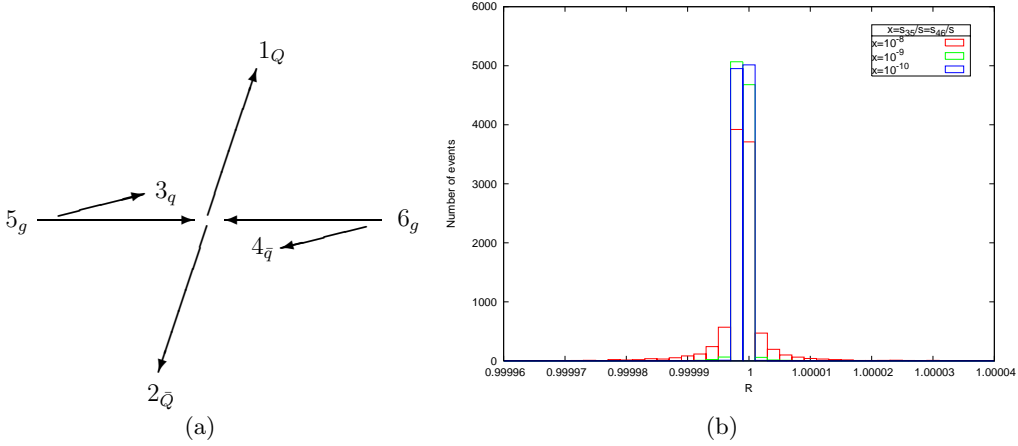}
\label{fig.dc}
}
\caption[]{\subref{fig.pic1} Ilustration of a double collinear event. \subref{fig.ds} Distribution of R for 10000 double collinear phase space points.}
\end{figure}
In fig.~\ref{fig.pic3} we represent schematically the kinematical configuration where $\gli{5}||\q{3}+\gli{6}||\qb{4}$. As it can be seen in fig.~\ref{fig.dc}, the convergence of the subtraction term to the real radiation matrix elements is indeed achieved as we make the control variable smaller. The same results are obtained for the double collinear limit $\gli{6}||\q{3}+\gli{5}||\qb{4}$.

\subsection{Final-final single collinear limit}
In addition to the double unresolved limits discussed above, the double real radiation matrix element contains two types of single collinear limits.
\begin{figure}[ht]
\centering
\subfigure[]{
\scalebox{0.8}{
\setlength{\unitlength}{4144sp}

\begingroup\makeatletter\ifx\SetFigFont\undefined
\gdef\SetFigFont#1#2#3#4#5{
  \reset@font\fontsize{#1}{#2pt}
  \fontfamily{#3}\fontseries{#4}\fontshape{#5}
  \selectfont}
\fi\endgroup
\begin{picture}(2927,2370)(1576,-2536)
\thicklines
{\put(3021,-1323){\vector(-2,-3){630.615}}
}
{\put(4321,-1276){\vector(-1, 0){1170}}
}
{\put(3090,-1243){\vector( 1, 2){449}}
}
{\put(3078,-1322){\vector( 0,-1){1035}}
}
{\put(3026,-1235){\vector( 1, 3){319.700}}
}
{\put(1756,-1276){\vector( 1, 0){1170}}
}
\put(1350,-1325){\makebox(0,0)[lb]{\smash{{\SetFigFont{12}{14.4}{\rmdefault}{\mddefault}{\updefault}{$5_g$}
}}}}
\put(4260,-1321){\makebox(0,0)[lb]{\smash{{\SetFigFont{12}{14.4}{\rmdefault}{\mddefault}{\updefault}{$6_g$}
}}}}
\put(2840,-2536){\makebox(0,0)[lb]{\smash{{\SetFigFont{12}{14.4}{\rmdefault}{\mddefault}{\updefault}{$1_Q$}
}}}}
\put(2050,-2440){\makebox(0,0)[lb]{\smash{{\SetFigFont{12}{14.4}{\rmdefault}{\mddefault}{\updefault}{$2_{\bar{Q}}$}
}}}}
\put(3110,-220){\makebox(0,0)[lb]{\smash{{\SetFigFont{12}{14.4}{\rmdefault}{\mddefault}{\updefault}{$3_{q}\:\:4_{\bar{q}}$}
}}}}

\end{picture}
\label{fig.pic4}
}}
\subfigure[]{
\includegraphics[scale=0.6]{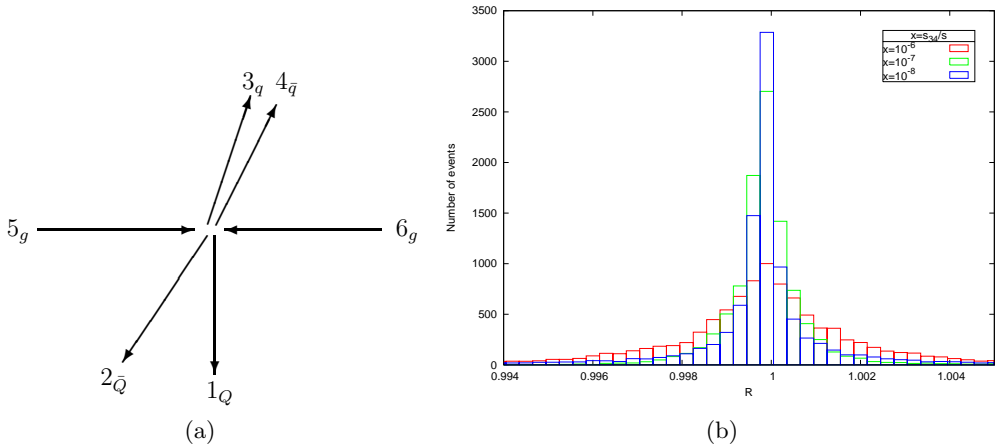}
\label{fig.scff}
}
\caption[]{\subref{fig.pic1} Ilustration of a final-final single collinear event. \subref{fig.ds} Distribution of R for 10000 final-final single collinear phase space points.}
\end{figure}
For the final-final collinear limit $\q{3}||\qb{4}$ depicted in fig.\ref{fig.pic4} we obtain the results shown in fig.\ref{fig.scff}, where, once more, a good convergence of the subtraction terms to the real radiation contributions is achieved as the limit is approached. It should be noted that due to the presence of angular correlation terms this good convergence is only achieved after integration over the azimuthal angle of the collinear pair. The procedure by which these angular correlations are integrated out has been explained in the case where massive partons are present  in~\cite{Abelof:2011ap}.

\subsection{Initial-final single collinear limits}
The last type of unresolved limits are the four collinear limits between the massless final state (anti) quark and one of the initial state gluons.
\begin{figure}[ht]
\centering
\subfigure[]{
\scalebox{0.8}{
\setlength{\unitlength}{4144sp}

\begingroup\makeatletter\ifx\SetFigFont\undefined
\gdef\SetFigFont#1#2#3#4#5{
  \reset@font\fontsize{#1}{#2pt}
  \fontfamily{#3}\fontseries{#4}\fontshape{#5}
  \selectfont}
\fi\endgroup
\begin{picture}(2927,2640)(1576,-2581)

\thicklines
{\put(3021,-1323){\vector(-2,-3){640}}
}
{\put(1756,-1276){\vector( 1, 0){1170}}
}
{\put(4321,-1276){\vector(-1, 0){1170}}
}
{\put(3026,-1235){\vector( 1, 3){364.500}}
}
{\put(1910,-1243){\vector( 4, 1){641.412}}
}
{\put(3100,-1327){\vector(0,-1){1152.6}}
}
\put(1340,-1330){\makebox(0,0)[lb]{\smash{{\SetFigFont{12}{14.4}{\rmdefault}{\mddefault}{\updefault}{$5_g$}
}}}}
\put(4270,-1321){\makebox(0,0)[lb]{\smash{{\SetFigFont{12}{14.4}{\rmdefault}{\mddefault}{\updefault}{$6_g$}
}}}}
\put(2460,-1110){\makebox(0,0)[lb]{\smash{{\SetFigFont{12}{14.4}{\rmdefault}{\mddefault}{\updefault}{$3_q$}
}}}}
\put(2070,-2500){\makebox(0,0)[lb]{\smash{{\SetFigFont{12}{14.4}{\rmdefault}{\mddefault}{\updefault}{$2_{\bar{Q}}$}
}}}}
\put(3230,-61){\makebox(0,0)[lb]{\smash{{\SetFigFont{12}{14.4}{\rmdefault}{\mddefault}{\updefault}{$1_Q$}
}}}}
\put(2830,-2670){\makebox(0,0)[lb]{\smash{{\SetFigFont{12}{14.4}{\rmdefault}{\mddefault}{\updefault}{$4_{\bar{q}}$}
}}}}

\end{picture}
\label{fig.pic5}
}}
\subfigure[]{
\includegraphics[scale=0.6]{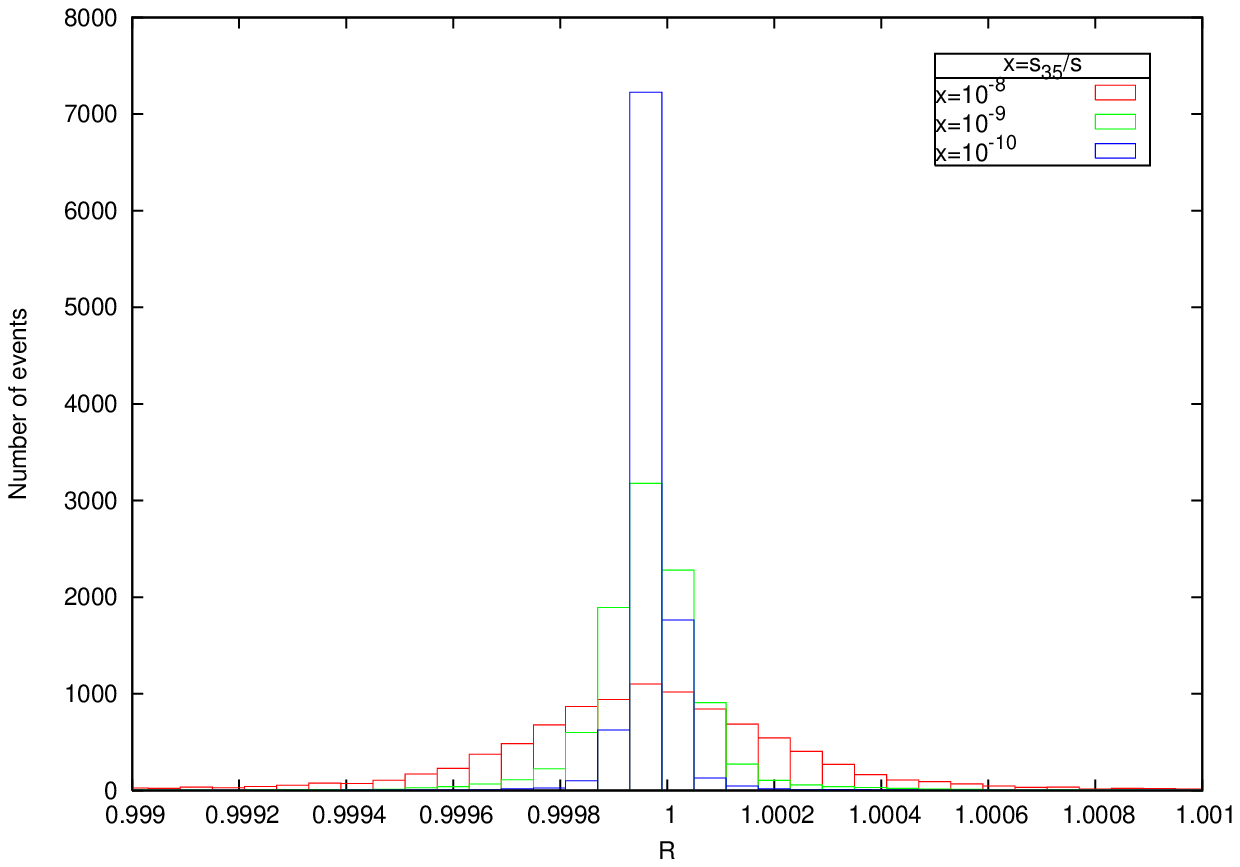}
\label{fig.scif}
}
\caption[]{\subref{fig.pic1} Ilustration of an initial-final single collinear event. \subref{fig.ds} Distribution of R for 10000 initial-final single collinear phase space points.}
\end{figure}
As it can be seen in fig.\ref{fig.scif}, also in these limits our subtraction terms constitute a valid approximation of the real radiation differential cross section. As it is indicated in fig.\ref{fig.pic5} the histogram in fig.\ref{fig.scif} corresponds to the limit $\gli{5}||\q{3}$. Similar results are also obtained for the other three initial-final single collinear limits: $\gli{6}||\q{3}$, $\gli{5}||\qb{4}$, and $\gli{6}||\qb{4}$.

\section{Conclusions}\label{sec.conclusions}
In this paper, we present the double real radiation contributions to the $t \bar t$ hadronic production cross section coming from the partonic process $gg\rightarrow t\bar{t}q\bar{q}$. These contributions develop infrared divergencies when one or two partons become unresolved (soft or collinear) such that a systematic subtraction procedure is required before these contributions can be evaluated numerically.

We follow the formalism developed in \cite{Abelof:2011ap}, where we extended the NNLO antenna subtraction formalism to include the evaluation of hadronic observables involving a massive pair of fermions. Section 2 describes the infrared structure of double real contributions for these observables.

To capture the double unresolved singular behaviour of the gluon-gluon initiated matrix elements, we derive the appropriate four-parton antenna functions and establish their limiting behaviour respectively in sections \ref{sec.antennae} and \ref{sec.limits}. Using these, in section~\ref{sec.subterms}, we explicitly construct the antenna subtraction terms for the $gg\to t\bar t q\bar q$ subprocess in leading and subleading colour contributions.

In section \ref{sec.results}, we check numerically that our subtraction terms approximate the real matrix elements in all single and double unresolved configurations in a point-by point manner. In the regions of phase space associated to the single unresolved collinear kinematical configurations, the ratio between real matrix elements and subtraction term approaches unity, provided the azimuthal terms associated with these collinear limits are correctly treated.

The double real corrections stemming from the partonic channel $gg\rightarrow t\bar{t}q\bar{q}$ and contributing to the hadronic production of a top-antitop pair presented here provide a substantial step towards the calculation of the NNLO corrections to the top-quark pair production at the LHC. Future steps include in particular the computation of the remaining double real subtraction terms related to partonic channels involving only gluons in initial and final state and the computation of mixed real-virtual contributions for all partonic channels involved.

\section{Acknowledgements}
We would like to thank Joao Pires and Oliver Dekkers for many useful discussions. This research was supported by the Swiss National Science Foundation (SNF) under contract PP00P2-139192 and in part by the European Commission through the 'LHCPhenoNet' Initial Training Network PITN-GA-2010-264564', which are hereby acknowledged.

\appendix
\section{Single unresolved factors}\label{sec.singlefactors}
In this appendix we list and very briefly explain the different single unresolved factors that appear throughout our treatment of the partonic processes $gg\to Q\bar{Q}q\bar{q}$. 

When two final-state massless particles $i$ and $j$ become collinear the following kinematics occur:
\beq\label{eq:zdefinal}
p_i\to z p_{ij},\quad p_j\to (1-z) p_{ij} ,\quad s_{ik}\to z s_{ijk},
\quad s_{jk}\to (1-z) s_{ijk}\,,
\eeq
The corresponding splitting functions in the conventional dimensional regularisation scheme are given by~\cite{Altarelli:1977zs}:
\beqa
&& P_{qg\rightarrow Q}(z)=\frac{1+(1-z)^2-\epsilon z^2}{z}\label{eq.splitting1}\\
&& P_{q\bar{q}\rightarrow G}(z)=\frac{z^2+(1-z)^2-\epsilon}{1-\epsilon}\label{eq.splitting2}\\
&& P_{gg\rightarrow G}(z)=2\left[\frac{z}{1-z}+\frac{1-z}{z}+z(1-z)\right].\label{eq.splitting3}
\eeqa

When the collinearity arises between an initial  ($i$) and a final state parton ($j$), the kinematics are given by
\beq\label{eq:zdefinitial}
p_j\to z p_{i},\quad p_{ij}\to (1-z) p_{i} ,
\quad s_{ik}\to \frac{s_{ijk}}{1-z},\quad s_{jk}\to \frac{z s_{ijk}}{1-z}\;.
\eeq
and the splitting functions are~\cite{Altarelli:1977zs}:
\beqa
&& P_{gq\leftarrow Q}(z)=\frac{1+z^2-\epsilon(1-z)^2}{(1-\epsilon)(1-z)^2}
=\frac{1}{1-z}\frac{1}{1-\epsilon}P_{qg\rightarrow Q}(1-z)\label{eq.splitting4}\\
&& P_{qg\leftarrow Q}(z)=\frac{1+(1-z)^2-\epsilon z^2}{z(1-z)}
=\frac{1}{1-z}P_{qg\rightarrow Q}(z)\label{eq.splitting5}\\
&& P_{q\bar{q}\leftarrow G}(z)=\frac{z^2+(1-z)^2-\epsilon}{1-z}
=\frac{1-\epsilon}{1-z}P_{q\bar{q}\rightarrow G}(z)\label{eq.splitting6}\\
&& P_{gg\leftarrow G}(z)=\frac{2(1-z+z^2)^2}{z(1-z)^2}
=\frac{1}{1-z}P_{gg\rightarrow G}(z).\label{eq.splitting7}
\eeqa
The additional factors $(1-\epsilon)$ and $1/(1-\epsilon)$ account for the different number of polarizations of quark and gluons in the cases in which the particle entering the hard processes changes its type.

In all the splitting functions defined above, the label $q$ can stand for a massless quark or an antiquark since charge conjugation implies that  $P_{qg\rightarrow Q}=P_{\bar{q}g\rightarrow\bar{Q}}$ and $P_{qg\leftarrow Q}=P_{\bar{q}g\leftarrow\bar{Q}}$. The labels $Q$ and $G$ denote the parent parton of the two collinear partons, which is massless. 

When a gluon $j$ emitted between the hard radiators $i$ and $k$ becomes soft, the eikonal factor that factorises off the colour-ordered squared matrix element is
\beq\label{eq.eikonalmassive}
S_{ijk}(m_i,m_k)=\frac{2s_{ik}}{s_{ij}s_{jk}}-\frac{2m_i^2}{s_{ij}^2}-
\frac{2m_k^2}{s_{jk}^2},
\eeq
where $m_i$ and $m_k$ are the masses of the radiators.

\section{Three-parton antennae}\label{sec.3partonantennae}
In this appendix we list the three-parton antenna functions used in the subtraction terms of section \ref{sec.subterms} together with their single unresolved limits. The massive and massless antennae can be found in~\cite{Abelof:2011jv,GehrmannDeRidder:2009fz} and~\cite{Daleo:2006xa,GehrmannDeRidder:2005cm} respectively, together with their integrated forms. All corresponding single unresolved factors can be found in appendix \ref{sec.singlefactors}. The massive A-type flavour violating initial-final antenna initiated by a gluon, is new. It will be given below in unintegrated and integrated form together with its infrared limits.

In order to make the mass-dependence in the expressions of the antenna functions explicit, we use the same convention as in the paper and define our invariants as $s_{ij}=2 p_{i} \cdot p_{j}$.

\subsection{Massive final-final antennae}
Our subtraction terms use A and E-type massive final-final antennae. The former is given by
\beqa
A^{0}_{3}(\Q{1},\gl{3},\Qb{2})& =& \frac{1}{ \left(E_{cm}^2 + 2 m_{Q}^2\right)} \left(\frac{2 s_{12}^2}{s_{13} s_{23}}+\frac{2s_{12}}{s_{13}}+\frac{2s_{12}}{s_{23}}+\frac{s_{23}}{s_{13}}+\frac{s_{13}}{s_{23}}\right.\nonumber\\
&&\left. + m_{Q}^2 \left(\frac{8 s_{12}}{s_{13} s_{23}}-\frac{2s_{12}}{s_{13}^2}-\frac{2s_{12}}{s_{23}^2}-\frac{2s_{23}}{s_{13}^2}-\frac{2}{s_{13}}-\frac{2}{s_{23}}-\frac{2 s_{13}}{s_{23}^2}\right)\right.\nonumber\\
&&\left. + m_{Q}^4 \left(-\frac{8}{s_{23}^2}-\frac{8}{s_{13}^2}\right)\right)+\order{\epsilon},\label{eq:A03mff}
\eeqa
with $E_{cm}^2=(p_1+p_2+p_3)^2$. 
It has only a single soft limit:
\beq
A_3^0(\Q{1},\gl{3},\Qb{2})\stackrel{^{p_3 \rightarrow0}}{\longrightarrow} \ssoft{1}{3}{2}(m_Q,m_Q).
\eeq
This antenna has been derived and integrated in \cite{GehrmannDeRidder:2009fz}.

The E-type antenna is
\beq\label{eq.Effm}
 E^{0}_{3}(\Q{1},\q{3},\qb{4}) = \frac{1}{\left(E_{cm}^2-m_{Q}^2 \right)^2} \left( s_{13} + s_{14} + \frac{s_{13}^2}{s_{34}} +\frac{s_{14}^2}{s_{34}} - 2 E_{cm} m_{Q}\right) + \order{\epsilon},
\eeq
where $E_{cm}^2=(p_1+p_3+p_4)^2$. Its only infrared limit is
\beq
 E^{0}_{3}(\Q{1},\q{3},\qb{4})\stackrel{^{\q{3}||\qb{4}}}{\longrightarrow}\frac{1}{s_{34}}P_{q\bar{q}\rightarrow G}(z).
\eeq
This antenna has been derived and integrated in~\cite{Abelof:2011jv}.

\subsection{Massless initial-final antennae}
We use the following A-type antenna
\beq
a_3^0(\q{1},\gli{3},\qb{2})=\frac{1}{s_{123}\left(s_{13}+s_{23}\right)}\left(  -\frac{2 s_{12}^2}{s_{13}}+\frac{2 s_{23} s_{12}}{s_{13}}+2s_{12}-\frac{s_{23}^2}{s_{13}}-s_{23}\right)+\order{\epsilon},
\eeq
with $s_{123}=s_{12}-s_{13}-s_{23}$. It is obtained by partial fractionning the full antenna $A_3^0(\q{1},\gli{3},\qb{2})$ so that it only contains the following single collinear limit
\beq
a_3^0(\q{1},\gli{3},\qb{2})\stackrel{^{1_q||3_g}}{\longrightarrow}\frac{1}{s_{13}}P_{qg\leftarrow Q}(z).
\eeq

We also need the following G-type antenna
\beq
G_3^0(\gli{1},\q{3},\qb{4})=\frac{1}{s_{134}^2}\left( \frac{s_{13}^2}{s_{34}}+ \frac{s_{14}^2}{s_{34}}\right)+\order{\epsilon}
\eeq
which has the single collinear limit
\beq
G_3^0(\gli{1},\q{3},\qb{4})\stackrel{^{\q{3}||\qb{4}}}{\longrightarrow}\frac{1}{s_{34}}P_{q\bar{q}\rightarrow G}(z).
\eeq
All massless initial-final antennae have been presented in unintegrated and in integrated forms in~\cite{Daleo:2006xa}.

\subsection{Massive initial-final antennae}

\parindent 0em

{\bf{Initial-Final E and D-type antennae}}\\
The massive initial-final three-parton D-type antenna is given by
\beqa
D_3^0(\Q{1},\gl{3},\gli{4})&=&\frac{1}{\left(Q^2+m_Q^2 \right)^2}\bigg( 9 s_{13}-9 s_{14}-6 s_{34} +\frac{4s_{14}^2}{s_{13}}+\frac{s_{34}^2}{s_{13}}-\frac{s_{34}^2}{s_{14}}+\frac{3 s_{14} s_{34}}{s_{13}}\nonumber\\
&&+\frac{3s_{13} s_{34}}{s_{14}}-\frac{4s_{13}^2}{s_{14}}+\frac{2 s_{14}^3}{s_{13} s_{34}}-\frac{4s_{13}^2}{s_{34}}-\frac{4 s_{14}^2}{s_{34}}+\frac{6 s_{13}s_{14}}{s_{34}}+\frac{2 s_{13}^3}{s_{14} s_{34}}+2 m_Q m_{\chi}\nonumber\\
&&+m_Q^2 \left(-\frac{2s_{13}^2}{s_{14}^2}+\frac{4 s_{34} s_{13}}{s_{14}^2}+\frac{4s_{13}}{s_{14}}-\frac{2 s_{34}^2}{s_{14}^2}-\frac{6s_{34}}{s_{14}}+\frac{2 s_{34}^2}{s_{14} s_{13}}+\frac{4s_{14}}{s_{13}}\right.\nonumber\\
&&\left.+\frac{6 s_{34}}{s_{13}}-\frac{2s_{14}^2}{s_{13}^2}-\frac{2 s_{34}^2}{s_{13}^2}-\frac{4 s_{14}s_{34}}{s_{13}^2}-6\right)-\frac{2 m_Q^3 m_{\chi}s_{34}}{s_{13} s_{14}}+\frac{2 m^4 s_{34}}{s_{13} s_{14}}\bigg) +\order{\epsilon},\nonumber\\
\eeqa
where $Q^2=-(p_1+p_3-p_4)^2$ and $m_{\chi}=\sqrt{Q^2}$. It has a soft gluon limit as well as a single collinear limit
\beqa
&&D_3^0(\Q{1},\gl{3},\gli{4})\stackrel{^{p_3 \rightarrow0}}{\longrightarrow} \ssoft{1}{3}{4}(m_Q,0),\\
&&D_3^0(\Q{1},\gl{3},\gli{4})\stackrel{^{3_g||4_g}}{\longrightarrow}\frac{1}{s_{34}}P_{gg\leftarrow G}(z).
\eeqa
This antenna is related to the split antennae $D_3^0(\gl{4};\gl{3},\Q{1})$ and $D_3^0(\gl{4};\Q{1},\gl{3})$ defined in~\cite{Abelof:2011jv} through
\beq
D_3^0(\Q{1},\gl{3},\gli{4})=D_3^0(\gl{4};\gl{3},\Q{1})-D_3^0(\gl{4};\Q{1},\gl{3}).
\eeq\\

\parindent 1.5em

The massive initial-final E-type antenna used in our subtraction terms reads
\beq\label{eq.Eifm}
E_3^0(\Q{1},\q{3},\qi{4})=-\frac{1}{\left(Q^2+m_Q^2\right)^2}\left(-s_{14}+s_{13}-\frac{s_{13}^2}{s_{34}}-\frac{s_{14}^2}{s_{34}}
-2m_{Q}m_{\chi}\right)+\order{\epsilon},
\eeq
with $Q^2=-(p_1+p_3-p_4)^2$, and $m_{\chi}=\sqrt{Q^2}$. The only infrared limit that this antenna has is 
\beq
E_3^0(\Q{1},\q{3},\qi{4})\stackrel{^{3_q||\hat{4}_q}}{\longrightarrow}\frac{1}{s_{34}} P_{gq\leftarrow Q}(z).
\eeq
Both of these antennae have been derived and integrated in~\cite{Abelof:2011jv}.\\

\parindent 0em

{\bf{Massive flavour violating initial-final A-type antenna}}\\
The new three-parton  flavour violating  A-type antenna initiated by a gluon is denoted by $A_3^0(\Q{1},\gli{3},\qb{2})$ and reads
\beqa
A_3^0(\Q{1},\gli{3},\qb{2})&=&\frac{1}{\left( Q^2+m_Q^2 \right)}\left(\frac{2s_{12}^2}{s_{13}s_{23}}-\frac{2s_{12}}{s_{13}}-\frac{2s_{12}}{s_{23}}+\frac{s_{13}}{s_{23}}+\frac{s_{23}}{s_{13}}\right.\nonumber\\
&& \left. -m_Q^2\left(\frac{2s_{12}}{s_{13}^2}-\frac{2s_{23}}{s_{13}^2}-\frac{2}{s_{13}}\right)\right)+\order{\epsilon},\label{eq.Afl}
\eeqa
where $Q^2=-(p_1+p_2-p_3)^2$. 

\parindent 1.5em

Its infrared limit is
\beq
A_3^0(\Q{1},\gli{3},\qb{2})\stackrel{^{\qb{2}||\hat{3}_g}}{\longrightarrow}\frac{1}{s_{23}}P_{qg\leftarrow Q}(z).
\eeq

Its integrated form denoted by ${\cal A}_3^0(\Q{1},\gli{3},\qb{2})$ is obtained by integrating the expression given in eq.(\ref{eq.Afl}) over the unresolved
intial-final antenna phase space involving one massive parton of mass $m_{Q}$~\cite{Abelof:2011jv}. The integrated antenna  is given by, 
\begin{eqnarray}
{\cal A}_3^0(\Q{1},\gli{3},\qb{2})=\frac{1}{C(\epsilon)}
\int {\rm d}\Phi_2\frac{(Q^2+m_K^2)}{2\pi}A_3^0(\Q{1},\gli{3},\qb{2})
\end{eqnarray} 
where $C(\epsilon)$, is a normalisation factor given by
\beq\label{eq.Cepsilon}
C(\e)=(4 \pi)^{\e}\frac{e^{-\e\gamma_{E}}}{8 \pi^2}
\eeq
and the initial-final massive antenna phase space denoted by ${\rm d}\Phi_{X_{i,jk}}$ is given by,
\beq\label{eq.initialfinalphase}
{\rm d}\Phi_{X_{i,jk}}= {\rm d}\Phi_2\frac{(Q^2+m_K^2)}{2\pi}.
\eeq
In eq.(\ref{eq.initialfinalphase}), ${\rm d}\Phi_2$ is the massive $2\to 2$ phase space involving one massive final state particle, which we parametrise as~\cite{Abelof:2011jv} 
\beq\label{phasesp1m}
 {\rm d}\phi_2(m_Q,0)=\frac{(4\pi)^{\epsilon-1}}{2\Gamma(1-\epsilon)}
E_{cm}^{2\epsilon-2}\left( E_{cm}^2-m_Q^2 \right)^{1-2\epsilon}{\rm d}y\:y^{-\epsilon}(1-y)^{-\epsilon},
\eeq
where $y$ runs from 0 to 1. In this context, the invariants are given by,
\beqa
2p_i\cdot p_j &=&\frac{Q^2+m_Q^2}{xE_{cm}^2}\left[ E_{cm}^2-y\left( E_{cm}^2-m_Q^2\right)\right]\\
2p_i\cdot p_k &=&\frac{Q^2+m_Q^2}{xE_{cm}^2}\left[y\left( E_{cm}^2-m_Q^2\right)\right]\label{eq:sij1m}
\eeqa
and the center of mass energy $E_{cm}$ is
\beq
E_{cm}=\sqrt{\frac{Q^2(1-x)+m_Q^2}{x}}\hspace{1cm} {\rm with}\hspace{1cm}x=\frac{Q^2 +m_{j}^2}{2 p_{i}\cdot q}.
\eeq
The integrated antenna reads
\beqa
{\cal A}_3^0(\Q{1},\gli{3},\qb{2})&=& (Q^2+m_Q^2)^{-\e} \times \bigg[ - \frac{1}{2\e}p_{qg}^{(0)}(x) -\frac{1 - 4 x + 2 x^2 + x x_0 + 2 x^2 x_0 - 2 x^3 x_0}{2(1-x x_0)}  \nonumber\\
&&+(1-2x+2x^2)\left( \ln (1-x)-\frac{1}{2}\ln (x^2(1-x_0))\right)\bigg]+\order{\e}
\eeqa
with
\beq
x_0=\frac{Q^2}{Q^2+m_Q^2}
\eeq
and the splitting kernel $p_{qg}^{(0)}(x)$ given by
\beq
p_{qg}^{(0)}(x)=1-2x+2x^2.
\eeq
Having a single collinear limit in its unintegrated form, this integrated antenna develops a pole proportionnal to the splitting kernel  $p_{qg}^{(0)}(x)$, as expected.

\subsection{Massless initial-initial antennae}
Our subtraction terms employ two types of three-parton initial-initial antennae:  F and G-type antennae. The F-type antenna is given by
\beqa
F_3^0(\gli{1},\gl{2},\gli{3})&=&\frac{1}{s_{123}^2}\bigg(-12 s_{12}+12 s_{13}-12s_{23}+\frac{2 s_{12}^3}{s_{13} s_{23}}+\frac{4 s_{12}^2}{s_{13}}-\frac{4s_{12}^2}{s_{23}}\nonumber\\
&&+\frac{6 s_{23} s_{12}}{s_{13}}+\frac{6 s_{13}s_{12}}{s_{23}}+\frac{4 s_{23}^2}{s_{13}}-\frac{4 s_{13}^2}{s_{23}}+\frac{2 s_{23}^3}{s_{13}s_{12}}-\frac{4s_{13}^2}{s_{12}}\nonumber\\
&&-\frac{4s_{23}^2}{s_{12}}+\frac{6 s_{13} s_{23}}{s_{12}}+\frac{2s_{13}^3}{s_{23} s_{12}}\bigg)+\order{\epsilon}
\eeqa
with $s_{123}= s_{13} -s_{12}-s_{23}$. Its infrared limits are 
\beqa
&&F_3^0(\gli{1},\gl{2},\gli{3})\stackrel{^{p_2 \rightarrow0}}{\longrightarrow} \ssoft{1}{2}{3}(0,0),\\
&&F_3^0(\gli{1},\gl{2},\gli{3})\stackrel{^{\hat{1}_g||2_g}}{\longrightarrow}\frac{1}{s_{12}}P_{gg\leftarrow G}(z),\\
&&F_3^0(\gli{1},\gl{2},\gli{3})\stackrel{^{2_g||\hat{3}_g}}{\longrightarrow}\frac{1}{s_{23}}P_{gg\leftarrow G}(z).
\eeqa

Finally the initial-initial G-type antenna function is 
\beq
G_3^0(\gli{1},\q{3},\qi{4})=\frac{1}{s_{134}^2}\left( -\frac{s_{13}^2}{s_{34}}- \frac{s_{14}^2}{s_{34}}\right)+\order{\epsilon}.
\eeq
with, $s_{134}= s_{14} -s_{13}-s_{34}$.
It only has a single initial-final collinear limit
\beq
G_3^0(\gli{1},\q{3},\qi{4})\stackrel{^{3_q||\hat{4}_q}}{\longrightarrow}\frac{1}{s_{34}} P_{gq\leftarrow Q}(z).
\eeq
Both of these antennae have been derived and presented in their unintegrated and integrated forms in~\cite{Daleo:2006xa}.

\section{Four-parton antennae}\label{sec.4partonantennae}
In this section, we list the two known four-parton antenna functions used to construct our subtraction term presented in section \ref{sec.subterms}: the massive final-final B type antenna with a massive $Q \bar{Q}$ pair as radiators and the massless initial-initial G-type antenna with two initial state gluons as radiators. Their infrared limits were given in section \ref{sec.limits}.

\subsection{Massive final-final B-type antennae}
This antenna, denoted by $B_4^0(\Q{1},\qb{4},\q{3},\Qb{2})$, reads
\beqa
B_4^0(\Q{1},\qb{4},\q{3},\Qb{2})&=&\frac{1}{\left(E_{cm}^2+2m_Q^2\right)}\bigg\{ \frac{1}{s_{34}s_{134}^2}\left[ s_{12}s_{13}+s_{12}s_{14}+s_{13}s_{23}+s_{14}s_{24}\right]\nonumber\\
&&+\frac{1}{s_{34}s_{234}^2}\left[ s_{12}s_{23}+s_{12}s_{24}+s_{13}s_{23}+s_{14}s_{24}\right]\nonumber\\
&&+\frac{1}{s_{34}^2 s_{134}^2}\left[ 2s_{12}s_{13}s_{14}+s_{13}s_{14}s_{24}+s_{13}s_{14}s_{23}-s_{13}^2 s_{24}-s_{14}^2 s_{23}\right]\nonumber\\
&&+\frac{1}{s_{34}^2 s_{234}^2}\left[ 2s_{12}s_{23}s_{24}+s_{13}s_{23}s_{24}+s_{14}s_{23}s_{24}-s_{13} s_{24}^2-s_{14} s_{23}^2\right]\nonumber\\
&&+\frac{1}{s_{34}s_{134}s_{234}}\left[ 2s_{12}^2+s_{12}s_{23}+s_{12}s_{24}+s_{12}s_{13}+s_{12}s_{14}\right]\nonumber\\
&&+\frac{1}{s_{34}^2s_{134}s_{234}}\big[ -s_{13}s_{24}^2-s_{14}s_{23}^2-s_{13}^2 s_{24}-s_{14}^2 s_{23}\nonumber\\
&& +s_{13}s_{14}s_{23}+s_{13}s_{14}s_{24}+s_{13}s_{23}s_{24}+s_{14}s_{23}s_{24} \nonumber\\
&& -2s_{12}s_{13}s_{24}-2s_{12}s_{14}s_{23}\big]+\frac{2s_{12}}{s_{134}s_{234}}\nonumber\\
&& +m_Q^2\bigg[ \frac{8s_{13}s_{14}}{s_{34}^2 s_{134}^2}+\frac{8s_{23}s_{24}}{s_{34}^2 s_{234}^2}-\frac{4}{s_{134}^2}-\frac{4}{s_{234}^2}\nonumber\\
&&-\frac{2}{s_{34}s_{134}^2}\left[s_{12}+s_{23}+s_{24}\right]-\frac{2}{s_{34}s_{234}^2}\left[s_{12}+s_{13}+s_{14}\right]\nonumber\\
&&+\frac{2}{s_{34}s_{134}s_{234}}\left[4s_{12}-s_{13}-s_{14}-s_{23}-s_{24}\right]\nonumber\\
&& -\frac{8}{s_{34}^2 s_{134}s_{234}}\left[s_{14}s_{23}+s_{13}s_{24}\right]\bigg]\nonumber\\
&&-m_Q^4\bigg[\frac{8}{s_{34}s_{134}^2}+\frac{8}{s_{34}s_{234}^2}\bigg]\bigg\}+\order{\epsilon},\label{eq.B04ffm}
\eeqa
where $E_{cm}^2=(p_1+p_2+p_3+p_4)^2$. It is normalised to the tree-level two-parton matrix element (with couplings and colour factors omitted)
\beq
\left| \cm_2(\gamma^*\rightarrow Q\bar{Q})\right|^2=4\left[ (1-\epsilon)E_{cm}^2+2m_Q^2\right].
\eeq

\subsection {Massless initial-initial G-type antenna}
This antenna denoted by $G_4^0(\gli{1},\q{3},\qb{4},\gli{2})$ reads
\beqa
G_4^0(\gli{1},\q{3},\qb{4},\gli{2})&=& \frac{1}{Q^4} \Bigg\{ \frac{1}{s_{12}^2 s_{34}^2   } \left[ 2 s_{13} s_{14} s_{23} s_{24} -  s_{13}^2 s_{24}^2-  s_{14}^2 s_{23}^2\right]+ \frac{s_{23} }{s_{12} s_{13} s_{34}   } \left[s_{23}^2+  s_{24}^2\right]\nonumber \\ 
&&+ \frac{s_{23} }{s_{12} s_{13} s_{134}   } \left[s_{23}^2+2 s_{24} s_{34}+  s_{24}^2 +  s_{34}^2\right]+ \frac{s_{23} }{s_{12} s_{13}   } \left[s_{14} + 2 s_{24}+  s_{34} \right]\nonumber \\ 
&&+ \frac{s_{13}}{s_{12}    }- \frac{1}{s_{12} s_{34} s_{134}   } \Big[- 2 s_{14} s_{23}^2+ 2 s_{14} s_{24}^2+ 2 s_{14}^2 s_{23}+ 2 s_{14}^2 s_{24}+  s_{23} s_{24}^2\nonumber \\ 
&&+  s_{23}^2 s_{24}+  s_{23}^3+  s_{24}^3\Big]+ \frac{1}{s_{12} s_{34}   } \Big[2 s_{13} s_{14}+ 4 s_{13} s_{23}+ 3 s_{13} s_{24} + 2 s_{13}^2\nonumber \\ 
&&-  s_{14} s_{23}+ 4 s_{14}^2\Big]+ \frac{1}{s_{12} s_{134}   } \left[2 s_{14} s_{24} - 4 s_{23} s_{24} -  s_{23} s_{34}-  s_{24} s_{34}\right]\nonumber \\ 
&& + \frac{s_{12} }{s_{34} s_{134} s_{234}   } \left[-2 s_{12} s_{14}+  s_{12}^2+ 4 s_{14} s_{24}+ 4 s_{14}^2+ 4 s_{24}^2\right]\nonumber \\
&&+ \frac{s_{12}}{ s_{134} s_{234}   } \left[ -6 s_{14}- 6 s_{24}+ 3 s_{34}\right]+ \frac{2s_{12}^2 s_{14} s_{24}}{ s_{34}^2 s_{134} s_{234}   }+ \frac{1}{2s_{13} s_{24}   } \left[s_{12} s_{34}\right. \nonumber \\ 
&&\left.- s_{14} s_{23} \right]+ \frac{s_{24} }{s_{13} s_{34} s_{234}   } \left[ 2 s_{12} s_{24}+  s_{12}^2+ 2 s_{24}^2\right]- \frac{1}{s_{13} s_{34}   } \Big[ 2 s_{12} s_{23}\nonumber \\ 
&& - 2 s_{12} s_{24}-  s_{12}^2+ 2 s_{23} s_{24}  - 2 s_{23}^2- 2 s_{24}^2\Big]- \frac{s_{34} }{s_{13} s_{134}^2   } \Big[  2 s_{12} s_{23}+ 2 s_{12} s_{24}\nonumber \\ 
&&-  s_{12}^2- 2 s_{23} s_{24}-  s_{23}^2 -  s_{24}^2\Big]- \frac{1}{s_{13} s_{134} s_{234}   } \Big[-6 s_{12} s_{24} s_{34}+ 4 s_{12} s_{24}^2\nonumber \\ 
&&+ 3 s_{12} s_{34}^2+ 3 s_{12}^2 s_{24}- 3 s_{12}^2 s_{34} -  s_{12}^3+ 3 s_{24} s_{34}^2 - 4 s_{24}^2 s_{34} + 2 s_{24}^3 -  s_{34}^3 \Big]\nonumber \\ 
&&- \frac{1}{s_{13} s_{134}   } \Big[-8 s_{12} s_{23} + 4 s_{12} s_{24}- 4 s_{12} s_{34}+ 4 s_{12}^2+ 2 s_{23} s_{34}+ 6 s_{23}^2 + 2 s_{24}^2\nonumber \\ 
&&+ 2 s_{34}^2\Big]- \frac{1}{s_{13} s_{234}   } \left[ 4 s_{12} s_{24}- 2 s_{12} s_{34} +  s_{12}^2- 3 s_{24} s_{34}+ 4 s_{24}^2 +  s_{34}^2\right]\nonumber \\ 
&&- \frac{1}{s_{13}   } \left[ 2 s_{12}- 2 s_{23} + 2 s_{24}- 2 s_{34}\right]- \frac{s_{14}}{ s_{34}^2 s_{134}   } \Big[  4 s_{12} s_{14}- 8 s_{12} s_{24}+ 2 s_{12}^2\nonumber \\ 
&&+ 4 s_{14} s_{23} + 4 s_{14} s_{24}- 4 s_{23} s_{24}+ 4 s_{24}^2\Big] + \frac{s_{14}^2 }{s_{34}^2 s_{134}^2   } \Big[ 4 s_{12} s_{23}+ 4 s_{12} s_{24}\nonumber \\ 
&&- 2 s_{12}^2- 4 s_{23} s_{24}- 2 s_{23}^2- 2 s_{24}^2\Big] + \frac{1}{s_{34}^2   } \Big[  4 s_{12} s_{13} - 4 s_{12} s_{14}-  s_{12}^2\nonumber \\ 
&&+ 4 s_{13} s_{23}- 2 s_{13} s_{24}- 2 s_{13}^2+ 2 s_{14} s_{23}\Big]+ \frac{1}{s_{34} s_{134}   } \Big[-4 s_{12} s_{23}- 2 s_{12} s_{24} \nonumber \\
&&+ 5 s_{12}^2- 8 s_{14} s_{23}+ 6 s_{14} s_{24}+ 6 s_{14}^2+ 6 s_{23}^2+ 4 s_{24}^2\Big]+ \frac{1}{s_{34}   } \left[2 s_{12}- 2 s_{13}\right. \nonumber \\ 
&&\left.- 6 s_{14}\right]+ \frac{1}{s_{134}^2   } \left[ 2 s_{12} s_{23}+ 2 s_{12} s_{24}-  s_{12}^2 - 2 s_{23} s_{24}-  s_{23}^2  -  s_{24}^2\right]\nonumber \\ 
&&+ \frac{1}{s_{134}   } \left[ 4 s_{12} - 4 s_{14} + 4 s_{23} + 2 s_{24}+ 3 s_{34}\right]- \frac{7}{2}+\left(1\leftrightarrow 2\:,\:3 \leftrightarrow 4 \right)\bigg\}\nonumber\\
&&+\order{\epsilon},\nonumber\\ \label{eq.G04ii}
\eeqa
where $Q^2=s_{12}+s_{34}-s_{13}-s_{14}-s_{23}-s_{24}$, $s_{134}=s_{34}-s_{13}-s_{14}$, and $s_{234}=s_{34}-s_{23}-s_{24}$. This antenna is normalised to the matrix element squared associated to the process $gg \to H$ which is given by
\beq
\left| \cm_2(gg  \rightarrow H)\right|^2=\frac{1}{2}(1-\epsilon)Q^4.
\eeq

\bibliography{Main}

\end{document}